\documentclass[
reprint,
aps,
prd,        % or prc / prl
amsmath,
amssymb,
nofootinbib
]{revtex4-2}

\usepackage{graphicx}% Include figure files
\usepackage{dcolumn}% Align table columns on decimal point
\usepackage{bm}% bold math
\usepackage{hyperref}

\usepackage{multirow}
\hypersetup{
	colorlinks=true,
	linkcolor=blue,
	filecolor=blue,      
	urlcolor=blue,
	citecolor=blue
}

\begin{document}
	
	\title{
		Topological production of charmonia with event-shape engineering in
		\texorpdfstring{$pp$}{pp} collisions at
		\texorpdfstring{$\sqrt{s}=13$}{sqrt(s)=13} TeV using PYTHIA8
	}
	
	\author{Aswathy Menon Kavumpadikkal Radhakrishnan$^1$}%\email[]{Aswathy.Menon@cern.ch}
	\author{Suraj Prasad$^1$}%\email[]{Suraj.Prasad@cern.ch}
	\author{Neelkamal Mallick$^{1,2}$}%\email[]{Neelkamal.Mallick@cern.ch}
	\author{Raghunath Sahoo$^1$}\email[Corresponding author:]{Raghunath.Sahoo@cern.ch}
	
	\affiliation{$^1$Department of Physics, Indian Institute of Technology Indore, Simrol, Indore 453552, India}
	\affiliation{$^2$Department of Physics, University of Jyv\"askyl\"a, P.O. Box 35, FI-40014, Jyv\"askyl\"a, Finland}
	
	\date{\today}
	
	\begin{abstract}
		
		The production of heavy quarks (charm and beauty) in high-energy hadronic and nuclear collisions provides an excellent testing ground for the theory of strong interactions and validates models based on quantum chromodynamics (QCD). In this work, prompt and nonprompt production of $\rm{J/}\psi$ in $pp$ collisions at $\sqrt{s}=13$ TeV are studied as a function of transverse spherocity using PYTHIA8. $\rm{J/}\psi$ is reconstructed via its electromagnetic decay to dielectrons and dimuons, in mid- and forward-rapidity, respectively. Transverse spherocity, an event shape observable, is used to distinguish hard QCD events from the softer, isotropic ones. In PYTHIA8, the production of $\rm{J/}\psi$ can be influenced by the average number of multiple parton interactions ($\langle N_{\rm mpi} \rangle$), owing to the underlying events (UE), which have a dominant contribution to particle production at lower transverse momentum. Since transverse spherocity is correlated to $\langle N_{\rm mpi} \rangle$, this can serve as an experimentally accessible tool for event selection to study the underlying QCD processes influencing the prompt and nonprompt $\rm{J/}\psi$ production. This study reveals the correlation between heavy-flavor production dynamics and topological event selection in $pp$ collisions using PYTHIA8, whose relevance awaits experimental validation.
		
	\end{abstract}
	
	\maketitle
	
	\section{Introduction}
	\label{sec1}
	
	Experimental explorations at the world's most powerful particle accelerator facilities like the Large Hadron Collider (LHC) at CERN, Switzerland, and the Relativistic Heavy Ion Collider (RHIC) at BNL, USA aim to study the formation of a deconfined state of hot and dense sub-nuclear matter, called as Quark-Gluon Plasma (QGP), and also to understand the fundamental processes of ``hadronization", which explain the formation of hadrons out of quarks~\cite{Bass:1998vz}. During such violent collisions of nuclear matter, quarks and gluons are liberated as the relevant degrees of freedom for a very short duration ($\simeq 10^{-23}$ s). In this environment, heavy quarks (charm and beauty) are produced predominantly in the initial hard-partonic interactions, which can be reliably understood via perturbative quantum chromodynamics (pQCD). As heavy quarks are created early in the collision, their production provides an excellent testing ground for the theory of strong interaction \textit{i.e.} QCD. Within the framework of the QCD factorization theorem, the heavy quark production cross-section is expressed as the convolution of three terms: (i) the parton distribution functions (PDFs), (ii) the parton hard-scattering cross sections, and (iii) the heavy-flavor fragmentation functions~\cite{Collins:1989gx}. 
	
	Among the various hadrons containing heavy quarks, quarkonia---bound states of a heavy quark and its antiquark---have been studied extensively, with particular emphasis on the lightest charmonium vector meson, $\rm{J}/\psi$ ($c\bar{c}$). Although the initial production of $c\bar{c}$ pairs from hard-partonic interactions is fairly described by pQCD, the subsequent formation of a color-neutral bound state involves soft-scale, non-perturbative processes that are still not fully understood and are typically addressed via phenomenological models~\cite{Brambilla:2010cs}.

	The inclusive $\rm{J}/\psi$ production occurs via two distinct mechanisms, \textit{i.e.} prompt and non-prompt. Prompt $\rm{J}/\psi$ is produced directly in the hard scattering or formed from the radiative decay of higher charmonium states. They reflect the early collision dynamics and serve as ideal probes to test strong-interaction models and study suppression/regeneration in the QGP. In contrast, non-prompt $\rm{J}/\psi$ are the weak decay products of beauty hadrons~\cite{ALICE:2015nvt, ALICE:2018szk, CMS:2017exb}, which carry most of the momentum of their parent hadrons, thus providing insight into beauty production in elementary, hadronic, and nuclear collisions. Together, they offer complementary probes of quarkonium production mechanisms. These two production modes of $\rm{J}/\psi$ exhibit distinctive topological signatures: the decay vertex of non-prompt $\rm{J}/\psi$ is significantly displaced from the primary vertex compared to prompt $\rm{J}/\psi$ because of the longer lifetime of the weakly decaying parent beauty hadrons~\cite{CDF:2004jtw, Prasad:2023zdd}. Exploiting these features allows experimental separation of prompt and non-prompt contributions to inclusive production, which in turn provides crucial insights into hadronization mechanisms in both the charm and beauty sectors.

	In experiments, the inclusive $\rm{J}/\psi$ measurements are performed by reconstructing $\rm{J}/\psi$ through its electromagnetic decays to dileptons, as the hadronic decays are mostly suppressed, which is described by the Okubo–Zweig–Iizuka (OZI) rule. Among the experiments at the LHC, ALICE is capable of measuring the topological production of $\rm{J}/\psi$ at both midrapidity and forward rapidity regions~\cite{ALICE:2012pet}. The production of $\rm{J}/\psi$ at midrapidity ($|y| < 0.9$) is measured via the dielectron decay channel \textit{i.e.} $\rm{J}/\psi$ $\rightarrow$ $e^{+} + e^{-}$. In contrast, the production at the forward rapidity ($2.5 < y < 4$) is measured via dimuon decay channel \textit{i.e.} $\rm{J}/\psi$ $\rightarrow$ $\mu^{+} + \mu^{-}$.

	In Refs.~\cite{ALICE:2012pet, ALICE:2020msa, ALICE:2021zkd, ALICE:2025fzz}, the ALICE Collaboration explores the production of self-normalized yield of inclusive J/$\psi$, as a function of self-normalized charged particle density at midrapidity for event selection based on multiplicity at both mid-rapidity and forward rapidity regions. Here, the collision energy dependence of the self-normalized yield of J/$\psi$ is also explored. Such correlation studies of charmonium production with charged particle multiplicity ($N_{\rm ch}$) are important to explore the interplay between hard and soft mechanisms of $\rm{J}/\psi$ production, both at the partonic level and at hadronization~\cite{Weber:2018ddv}. One observes that the self-normalized yield of J/$\psi$ at forward rapidity increases linearly as a function of self-normalized charged particle density at midrapidity when event classification is performed based on multiplicity at midrapidity in $pp$ collisions at $\sqrt{s}=5.02$ TeV~\cite{ALICE:2021zkd}. This follows the $x=y$ line. However, the self-normalized yield of J/$\psi$ starts to deviate from the linear behaviour with an increase in $\sqrt{s}$. The deviations from the linear behaviour become stronger when the measurement of J/$\psi$ is changed from forward rapidity to midrapidity regions, irrespective of event classification based on midrapidity or forward rapidity multiplicity~\cite{ALICE:2020msa}. This observation underlines the fact that the production of J/$\psi$ is strongly affected when one moves from the midrapidity to forward rapidity regions. The phenomenological studies suggest that this non-linear behaviour is partially contributed from the non-prompt production of J/$\psi$~\cite{Weber:2018ddv}. In addition, the latest ALICE experimental data on the multiplicity-dependence of $\rm{J}/\psi$ where both $N_{\rm ch}$ and $\rm{J}/\psi$ yields are measured at forward rapidities, reiterate the finding of the steeper-than-linear increase of their correlation, attributing it to the autocorrelation effects~\cite{ALICE:2025fzz}. 
	Thus, it is necessary to make a systematic study of the prompt and non-prompt production of J/$\psi$ as a function of pseudorapidity.

	The pQCD-based event generators, such as PYTHIA8, are able to explain the heavy-flavour production up to a greater extent, in addition to a number of heavy-ion-like observations in high multiplicity $pp$ collisions~\cite{Prasad:2024gqq, Prasad:2023zdd, Goswami:2024xrx, Prasad:2025yfj}. This is possible thanks to the improved implementation of multipartonic interactions (MPI), color reconnection (CR), and rope hadronization (RH)~\cite{Prasad:2025yfj, Prasad:2024gqq}. The strange and multi-strange hadron ratios showing strangeness enhancement, as well as the particle ratios carrying information about the radial flow, are well explained by PYTHIA8 with MPI, CR, and RH~\cite{ALICE:2023bga, Prasad:2025yfj, Prasad:2024gqq}. Even the increase in charmonium production with charged-particle multiplicity is well explained by PYTHIA8 when MPI is included~\cite{ALICE:2015ikl, Thakur:2017kpv, Deb:2018qsl}. In fact, it is identified that the $\rm{J}/\psi$ production in $pp$ collisions is either directly linked to a strong hadronic activity, which increases particle production and biases the $dN_{\rm ch}/d\eta$ distributions to higher values, or the scenario is that the influence of MPI extends to the domain of hard processes, affecting the harder momentum scales relevant for quarkonia production~\cite{ALICE:2012pet}. Moreover, the production of high-momentum and heavier particles in the initial scattering restricts the energy available for MPI and, in turn, affects the particle produced through MPI and the event shape. However, estimating the number of MPI ($N_{\rm mpi}$) in experiments is not trivial. This challenge could be overcome if event-shape observables proportional to $N_{\rm mpi}$ are employed, one such event classifier being the transverse spherocity, $S_{0}$~\cite{Prasad:2025yfj}. In this context, the present work bridges this gap by utilising an event-shape–based approach--through transverse spherocity--to investigate the dynamics of ${\rm J}/\psi$ production.
	
	In small collision systems, event activity is often characterized using the charged-particle multiplicity measured at midrapidity or forward rapidity regions. While this approach is straightforward, it can suffer from autocorrelation biases, as the same particles may simultaneously define the event class and contribute to the observables being measured~\cite{Prasad:2025yfj, Prasad:2024gqq}. Such autocorrelations can artificially enhance apparent effects, making it difficult to disentangle genuine physics effects from methodological artifacts\footnote{For detailed discussions on the effects of autocorrelation, see~Refs.\cite{ALICE:2025fzz,Weber:2018ddv, ALICE:2020swj}.}. Although forward-rapidity-based multiplicity estimators reduce these biases compared to mid-rapidity selections, they do not eliminate them entirely. Event-shape observables offer a more differential and less biased way, compared to event classifiers based solely on multiplicity, to characterise the underlying event~\cite{ALICE:2019dfi}. They therefore provide a useful alternative for studying the interplay between soft and hard particle production with reduced sensitivity to autocorrelation effects~\cite{ALICE:2025fzz}. $S_0$ is one such event shape classifier, which enables us to segregate hard jetty-like events from soft-processes dominated isotropic events based on the geometry of particle production in each event, the classification strength of which is retained from $pp$ to Pb-Pb collisions~\cite{Mallick:2021hcs, Mallick:2020ium, Prasad:2025ezg, Tripathy:2025npe, Prasad:2022zbr, Prasad:2021bdq}. Interestingly, the classification strength of transverse spherocity changes weakly with a change in rapidity~\cite{MenonKavumpadikkalRadhakrishnan:2023cik}, which makes transverse spherocity a suitable event classifier to study the production of prompt and non-prompt J/$\psi$ in midrapidity and forward rapidity regions. Moreover, in addition to multiplicity-based studies, classifying collisions based on $S_{0}$ gives us the added benefit of analysing various mechanisms of particle production or hadronization for charm and beauty quarks~\cite{ALICE:2019dfi}. Thus, this paper explores the transverse spherocity and multiplicity-dependent production of prompt and non-prompt J/$\psi$ at midrapidity and forward rapidity regions in $pp$ collisions at $\sqrt{s}=13$ TeV using PYTHIA8 through several observables believed to be sensitive to the underlying production mechanisms and autocorrelation bias effects.
	J/$\psi$ production is studied via the electronic decay channel at midrapidity and the muonic decay channel at the forward rapidity regions. We focus on the observables such as partonic modification factor $Q_{\rm pp}$, mean transverse momentum $\langle p_{\rm T}\rangle$, non-prompt J/$\psi$ fraction $f_{\rm B}$, and self-normalized yield of prompt and non-prompt J/$\psi$.
	
	The paper is organised as follows. In Section~\ref{sec1}, we provide a brief introduction and motivation for the study. Section~\ref{sec2} discusses the event generation and methodology. In Section~\ref{sec:results}, we present our results and discuss the same. Finally, findings of the paper are summarised in Section~\ref{sec:summary}, with a brief future outlook.
	
	\section{Event generation and methodology}
	\label{sec2}
	In this section, we briefly discuss the event generation using PYTHIA8, along with the tunes used in this study. A brief discussion on the event-shape selection using transverse spherocity is also made.
	
	\subsection{PYTHIA8}
	\label{sec:PYTHIA8}
	The pQCD-based Monte Carlo event generators such as PYTHIA8~\cite{pythiamanual, Bierlich:2022pfr} describe many features of relativistic $pp$ collisions at the RHIC and LHC energies. PYTHIA8 has several models which incorporate soft and hard partonic interactions, initial and final state parton showers~\cite{Sjostrand:2004ef, Corke:2010yf}, multiple-partonic interactions (MPI)~\cite{Sjostrand:1987su}, beam remnants~\cite{Sjostrand:2004pf}, string fragmentation~\cite{Andersson:1983ia, Sjostrand:1984ic}, and hadronic decays~\cite{Andersson:1983ia, STAR:2010avo}. Here, along with MPI-based interactions, $2\to2$ hard processes are implemented, which leads to the production of charm and beauty hadrons.
	In this study, we have used the 4C-tune of PYTHIA8~\cite{Corke:2010yf} (version 8.308)\footnote{The results for 4C tune are compared with CR-BLC 2~\cite{Christiansen:2015yqa} in the Appendix~\ref{apxmodel}.}  to generate the minimum bias $pp$ collisions at $\sqrt{s}=13$ TeV, which include the inelastic and non-diffractive components (HardQCD:all = on) of the total collision cross section. In addition, a $p_{\rm T}$ cut-off of  $p_{\rm T} > 0.5$ GeV/c (using PhaseSpace:pTHatMinDiverge) is set to avoid the divergence of QCD processes in the limit $p_{\rm T} \rightarrow 0$. Along with the color reconnection (CR) and MPI, we have enabled all possible charmonia and bottomonia production processes available in PYTHIA using ``Charmonium:all=on" and ``Bottomonium:all=on". $\rm{J}/\psi$ is allowed to decay into opposite charge pairs of electrons and muons individually to have better statistics. For both electronic and muonic decay channels of $\rm{J}/\psi$, we produce 10 billion minimum bias events with the tunes mentioned above in $pp$ collisions at $\sqrt{s}=13$ TeV. A detailed description of event generation, along with identification of prompt and non-prompt $\rm{J}/\psi$, can be found in Ref.~\cite{Prasad:2023zdd}.

	In PYTHIA8, the production of heavy quark ($Q$) and anti-quark ($\bar{Q}$) pairs is implemented through the pQCD scattering processes which include the pair creation through gluon fusion ($gg\rightarrow Q\bar{Q}$) and light quark($q$)-antiquark($\bar{q}$) pair annihilation ($q\bar{q}\rightarrow Q\bar{Q}$)~\cite{Schwaller:2015gea}. In addition to this, the production of heavy quarks can be possible through the gluon splitting of parton showers, i.e., $g\rightarrow Q\bar{Q}$. At sufficiently large energy transfer, the heavy quarks also contribute to the parton distribution function, leading to the production of heavy quarks via flavour excitation ($gQ\rightarrow gQ$)~\cite{Weber:2018ddv}. Similarly, the quarkonia ($c\bar{c}$ or $b\bar{b}$) production in PYTHIA8 can also be implemented via several mechanisms. The leading-order nonrelativistic QCD (NRQCD) channels of $Q\bar{Q}$ production via colour-singlet and colour-octet pre-resonant states are included in the pQCD processes~\cite{Weber:2018ddv, Schwaller:2015gea}. The cluster collapse mechanism also contributes to quarkonia production when a heavy quark gets connected to a corresponding heavy antiquark close in phase space to form a quarkonium bound state, during the hadronization stage~\cite{Norrbin:1998bw}. If charmonia ($c\bar{c}$) is produced from the weak decay of a beauty hadron, it is referred to as non-prompt charmonium production.

	\subsection{Transverse spherocity}
	Transverse spherocity ($S_{0}$) is an event shape observable that characterizes the geometry of an event based on the azimuthal distribution of the particles produced. It is defined for a unit vector $\hat{n} (n_{\rm T},0)$ in the transverse plane, which minimises the sum of cross-products in the bracket as shown below~\cite{Banfi:2010xy, Prasad:2025yfj, ALICE:2023bga}:
	
	\begin{equation}
		S_{0}=\frac{\pi^2}{4}\min_{\hat{n}}\Bigg(\frac{\sum_{i=1}^{N_{\rm ch}}|\vec{p}_{{\rm T}_{i}} \times \hat{n}|}{\sum_{i=1}^{N_{\rm ch}}|\vec{p}_{{\rm T}_{i}}|}\Bigg)^{2}
		\label{eq:spherodefn}
	\end{equation}
	
	Here, $\vec{p}_{{\rm T}_{i}}$ stands for the transverse momentum vector of $i^{\rm th}$ charged particle, in an event and the iteration to find the suitable $\hat{n}$ runs over all charged particles ($N_{\rm ch}$). The multiplication factor $\pi^{2}/4$ in Eq.~\eqref{eq:spherodefn} is to normalize the quantity $S_{0}$ to lie between 0 and 1. Only the charged tracks within $|\eta| < 0.8$ with $p_{\rm T} >$ 0.15 GeV/$c$ are considered for the calculation of $S_{0}$. In this study, $S_{0}$ is constructed by setting $|\vec{p}_{{\rm T}_{i}}|=1.0$ to reduce the charge-to-neutral bias, where a minimum constraint of 10 charged particles at $|\eta| < 0.8$ and $p_{\rm T} >$ 0.15 GeV/$c$ is applied for each collision so that the description of unweighted spherocity is meaningful~\cite{ALICE:2023bga}. The lower limit, $S_{0} \rightarrow 0$, corresponds to jetty events containing back-to-back jets, which are consequences of hard scatterings. On the other hand, $S_{0} \rightarrow 1$ depicts isotropic events where particles are produced uniformly in the azimuthal plane, resulting from softer interactions. Thus, qualitatively, transverse spherocity $S_{0}$ disentangles pQCD-dominated hard processes from soft-QCD processes, helping in understanding the underlying particle production mechanism\footnote{Elaborate discussion on transverse spherocity can be found in Ref.~\cite{Prasad:2025yfj}.}. In this work, 20\% of events having the lowest and highest values of $S_{0}$ are respectively referred to as jetty and isotropic events. The cuts in $S_{0}$ for jetty and isotropic events for various V0M multiplicity classes, which
are estimated in the forward pseudorapidity ranges $2.8<\eta<5.1$ and
$-3.7<\eta<-1.7$, are shown in Table~\ref{spherocutPYTHIA}. A comparison of the spherocity distribution from PYTHIA8 results (Tune 4C and CR-BLC~2)\footnote{See Appendix~\ref{apxmodel} for details on CR-BLC~2 PYTHIA8 settings.} to that of the experimental data from the ALICE collaboration, for the highest V0M multiplicity class (0--1\%), is presented in Fig.~\ref{fig:S0datacomp} (Appendix~\ref{apxdata}), where a good qualitative and quantitative similarity is observed between them.
	
	\begin{table}[ht!]
		\begin{tabular}{|c||c||c||c|c|}
			\hline
			\multirow{2}{*}{\bf{V0M Percentile}} & \multirow{2}{*}{\bf{$\langle dN_{\rm ch}/d\eta\rangle_{|\eta|<0.5}$}} & \multicolumn{2}{c|}{\bf{$S_{0}$ range}} \\\cline{3-4}
			& & \bf{Jetty}  & \bf{Isotropic}  \\ \hline\hline
			0 -- 1      &20.17     & 0--0.646  & 0.844--1       \\ \hline
			1 -- 5      &12.47    & 0--0.576  &  0.812--1      \\ \hline
			5 -- 10     & 7.81     & 0--0.528  &   0.784--1    \\ \hline
			10 -- 20    & 5.27     & 0--0.504  &   0.770--1      \\ \hline
			20 -- 40    & 3.34     & 0--0.486  &  0.760--1      \\ \hline
			40 -- 100    &  2.39     & 0--0.468  &  0.750--1      \\ \hline
			0 -- 100    & 3.74    & 0--0.525  & 0.785--1      \\ \hline
			
		\end{tabular}
		\caption{\label{spherocutPYTHIA} Transverse spherocity cuts for the jetty and isotropic events for different multiplicity classes in $pp$ collisions at $\sqrt{s}$ = 13 TeV using PYTHIA8(Tune 4C). The errors for the mean charged particle multiplicity density are too small ($<$~1~\%), hence not tabulated.} 
	\end{table}
	
	% %______________________________-
	% \begin{table}[ht!]
		% \begin{tabular}{|c||c||c||c|c|}
			% \hline
			% \multirow{2}{*}{\bf{V0M Percentile}} & \multirow{2}{*}{\bf{$\langle dN_{\rm ch}/d\eta\rangle_{|\eta|<0.5}$}} & \multicolumn{2}{c|}{\bf{$S_{0}$ range}} \\\cline{3-4}
			%                         & & \bf{Jetty}  & \bf{Isotropic}  \\ \hline\hline
			% 0 -- 1      &21.26     & 0--0.656  & 0.850--1       \\ \hline
			% 1 -- 5      &13.48    & 0--0.592  &  0.818--1      \\ \hline
			% 5 -- 10     & 8.86     & 0--0.546  &   0.794--1    \\ \hline
			% 10 -- 20    & 6.21    & 0--0.522  &   0.778--1      \\ \hline
			% 20 -- 40    & 4.22     & 0--0.504  &  0.768--1      \\ \hline
			% 40 -- 100    &  2.55     & 0--0.486  &  0.758--1      \\ \hline
			% 0 -- 100    & 4.25    & 0--0.535  & 0.795--1      \\ \hline
			
			% \end{tabular}
		% \caption{\label{spherocutPYTHIA-CRBLC} Transverse spherocity cuts for the jetty and isotropic events for different multiplicity classes in $pp$ collisions at $\sqrt{s}$ = 13 TeV using PYTHIA8 (Monash with CR-BLC tune). The errors for the mean charged particle multiplicity density are too small ($<$~1~\%), hence not tabulated.} 
		% \end{table}

	\section{Results and discussions}
	\label{sec:results}
	
	\begin{figure*}[ht!]
		\includegraphics[scale=0.44]{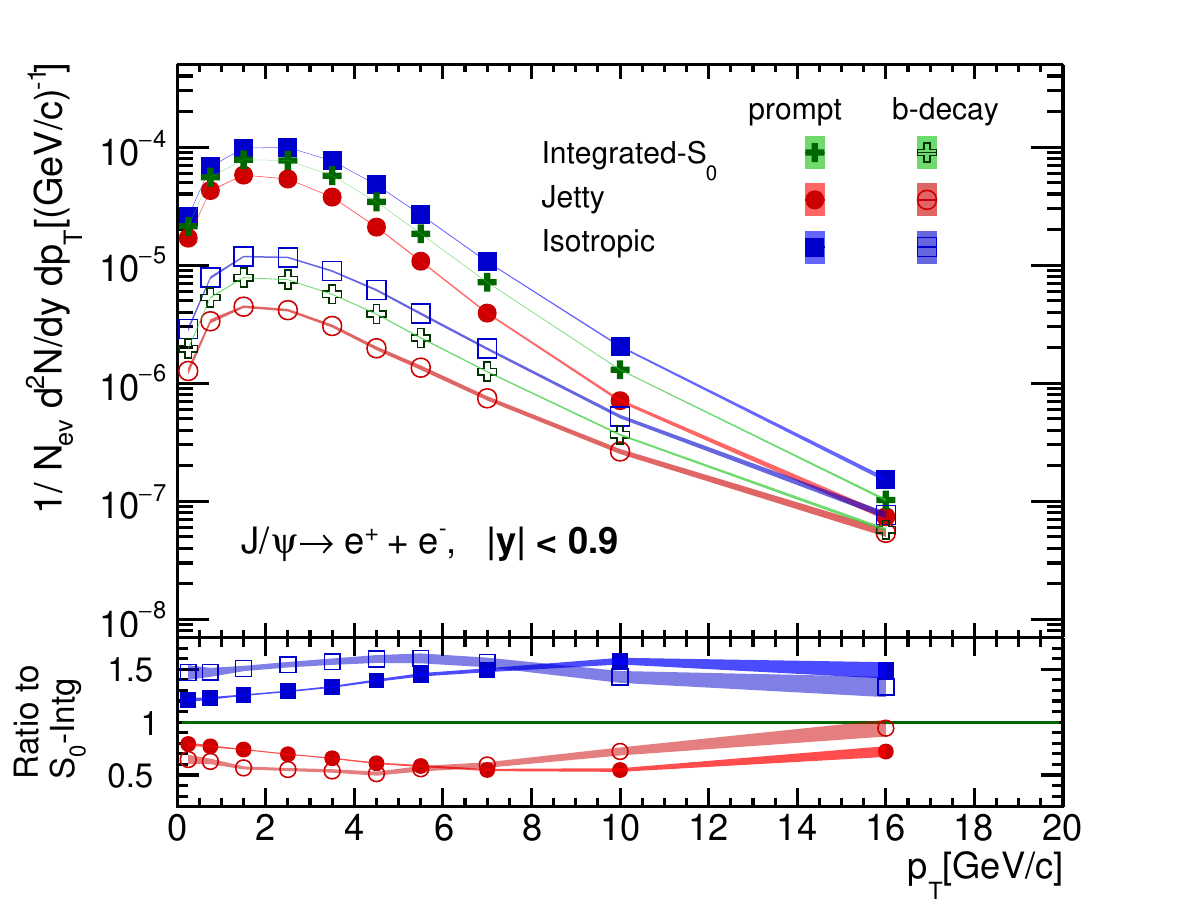}
		\includegraphics[scale=0.44]{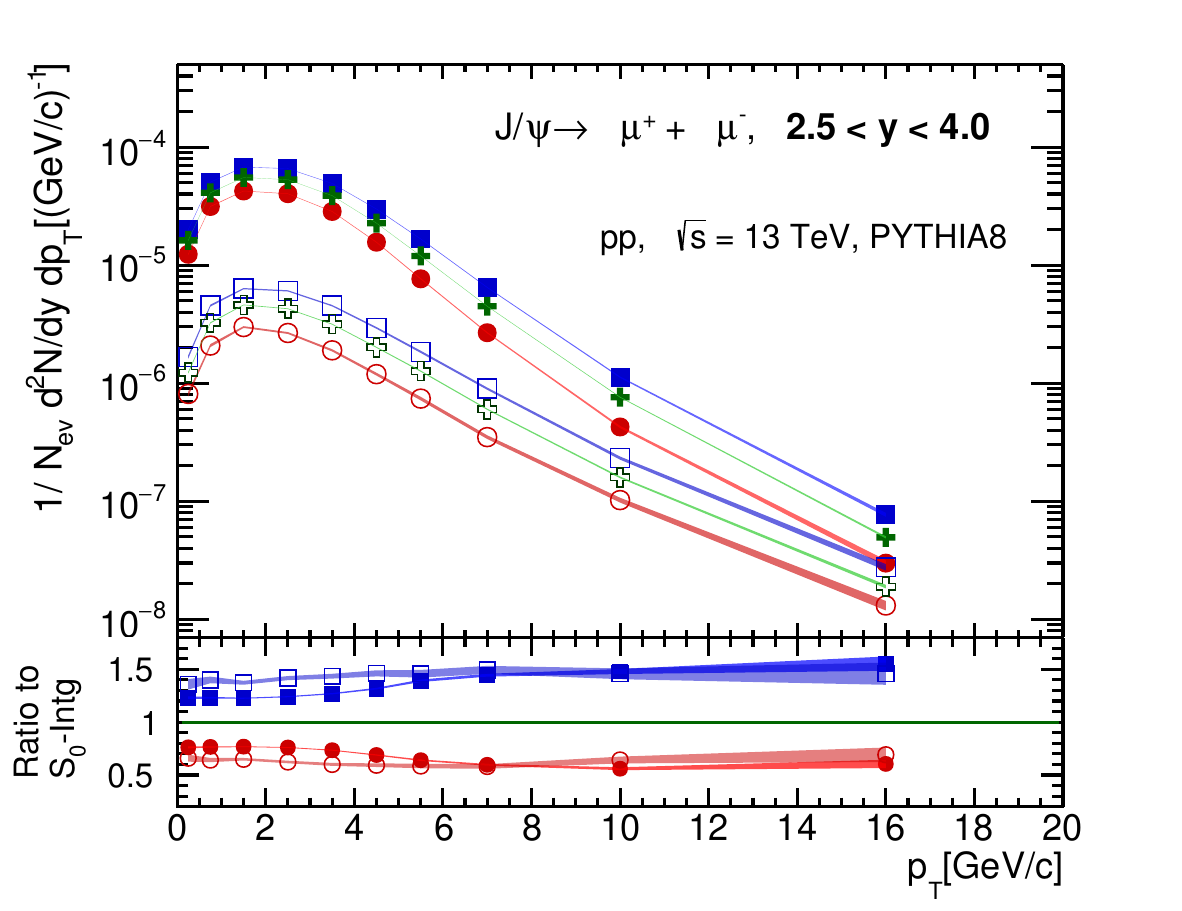}
		\caption{(Color online) Event-normalized transverse momentum ($p_{\rm T}$) spectra of prompt and non-prompt $\rm{J}/\psi$ in $S_{0}$-integrated, jetty and isotropic events from $pp$ collisions at $\sqrt{s}$ = 13 TeV using PYTHIA8, for dielectron (left) and dimuon (right) production channels at $|y| < 0.9$ and $2.5 < y < 4$ rapidity ranges respectively for (0--100\%) V0M class. The ratio of $p_{\rm T}$ spectra of the jetty and isotropic to $S_{0}$-integrated events is shown in the bottom panel. Error bands incorporate the estimated statistical uncertainties.}
		\label{fig:pTS0MUEch}
	\end{figure*}
	
	This section explains the results we obtained, as a function of transverse spherocity ($S_{0}$) for prompt and non-prompt $\rm{J}/\psi$ produced in $pp$ collisions at $\sqrt{s}$~=~13~TeV using PYTHIA8. Here, $\rm{J}/\psi$ is reconstructed via the dileptonic decay channels $\rm{J}/\psi$ $\rightarrow$ $e^{+} + e^{-}$ at midrapidity ($|y| < 0.9$) and $\rm{J}/\psi$ $\rightarrow$ $\mu^{+} + \mu^{-}$ in the forward rapidity ($2.5 < y < 4$). From this point onwards, $\rm{J}/\psi$ in the mid-rapidity and forward rapidity respectively implies the reconstruction using $\rm{J}/\psi$ $\rightarrow$ $e^{+} + e^{-}$ channel and $\rm{J}/\psi$ $\rightarrow$ $\mu^{+} + \mu^{-}$ channel.
	
	Shown in Fig.~\ref{fig:pTS0MUEch} are the $S_{0}$ dependence of event-normalized $p_{\rm T}$ spectra of prompt and non-prompt $\rm{J}/\psi$ produced in $pp$ collisions simulated using PYTHIA8 at $\sqrt{s}$~=~13~TeV for (0--100\%) V0M class. The $p_{\rm T}$ spectra of J/$\psi$, reconstructed via dielectron and dimuon channels, are shown in the left and right plots of Fig.~\ref{fig:pTS0MUEch}, respectively. We observe that the yield of non-prompt $\rm{J}/\psi$ is almost ten times smaller than that of prompt $\rm{J}/\psi$ in the low $p_{\rm T}$ regime while this difference shrinks as we approach higher values of $p_{\rm T}$ for both the rapidity regions, as the non-prompt $\rm{J}/\psi$ grows harder~\cite{Weber:2018ddv,ALICE:2012vpz}. Towards the low-$p_{\rm T}$ region, the contributions of prompt J/$\psi$ come primarily from the charmonia produced in the leading order semi-hard processes, i.e., $q\bar{q}\rightarrow C$ and $gg\rightarrow C$ (where $C$ refers to final state charmonium)~\cite{Deb:2018qsl}. Here, the probability of beauty hadron production is significantly small due to its large mass compared to charmonium states.
	Further, in Ref.~\cite{Weber:2018ddv}, the authors show that at high $p_{\rm T}$-regime, the particle production is largely dominated by the hard scatterings, where the production of beauty hadrons is comparable to charmonium states. Thus, the hardening of the $p_{\rm T}$-spectrum of non-prompt J/$\psi$ as compared to prompt  J/$\psi$ is a consequence of the competing effects of two different production mechanisms of charmonium states and beauty hadrons in different regions of transverse momentum. We note that these trends are in line with the results obtained from experiments~\cite{ALICE:2021edd, ATLAS:2018hqe, CMS:2010nis}.
	
	We also see a considerable dependence of transverse spherocity on the $p_{\rm T}$-dependent yield of topological production of J/$\psi$ in both the rapidity regions, as shown in Fig.~\ref{fig:pTS0MUEch}. Here, the production of both prompt and non-prompt $\rm{J}/\psi$ is observed to be favoured more in isotropic events as compared to the jetty events. In Refs.~\cite{MenonKavumpadikkalRadhakrishnan:2023cik, Prasad:2024gqq, Prasad:2025yfj}, it is explored that the transverse spherocity possesses a large correlation with $N_{\rm mpi}$. Additionally, it is also concluded in Ref.~\cite{Deb:2018qsl} that the production of inclusive J/$\psi$, which are mostly prompt in nature, is favored in events having large $N_{\rm mpi}$. In fact, the yield of inclusive J/$\psi$ increases linearly with $N_{\rm mpi}$~\cite{Deb:2018qsl}. The observation of a higher(lower) prompt J/$\psi$ yield in isotropic(jetty) events, which corresponds to events having large(small) $N_{\rm mpi}$ as compared to $S_{0}$-integrated events, is a testimony of the observations shown in Ref.~\cite{Deb:2018qsl}. Analogously, it can also be argued that the events having large $N_{\rm mpi}$ favor the $b$-hadron production, as compared to events having small $N_{\rm mpi}$, which results in a larger yield of non-prompt J/$\psi$ for isotropic events, as compared to the jetty events~\cite{Weber:2018ddv}. 
	
	\begin{figure*}[ht!]
		\includegraphics[scale=0.42]{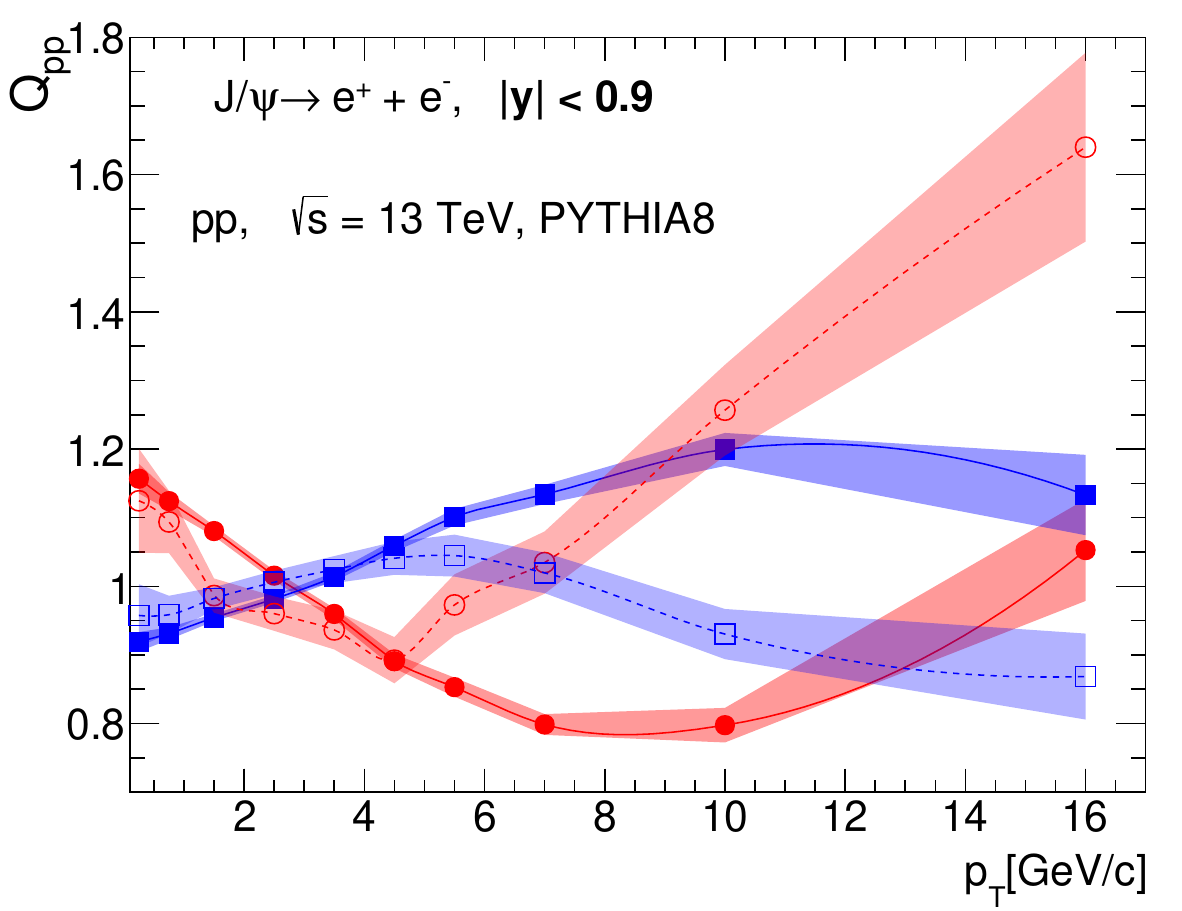}
		\includegraphics[scale=0.42]{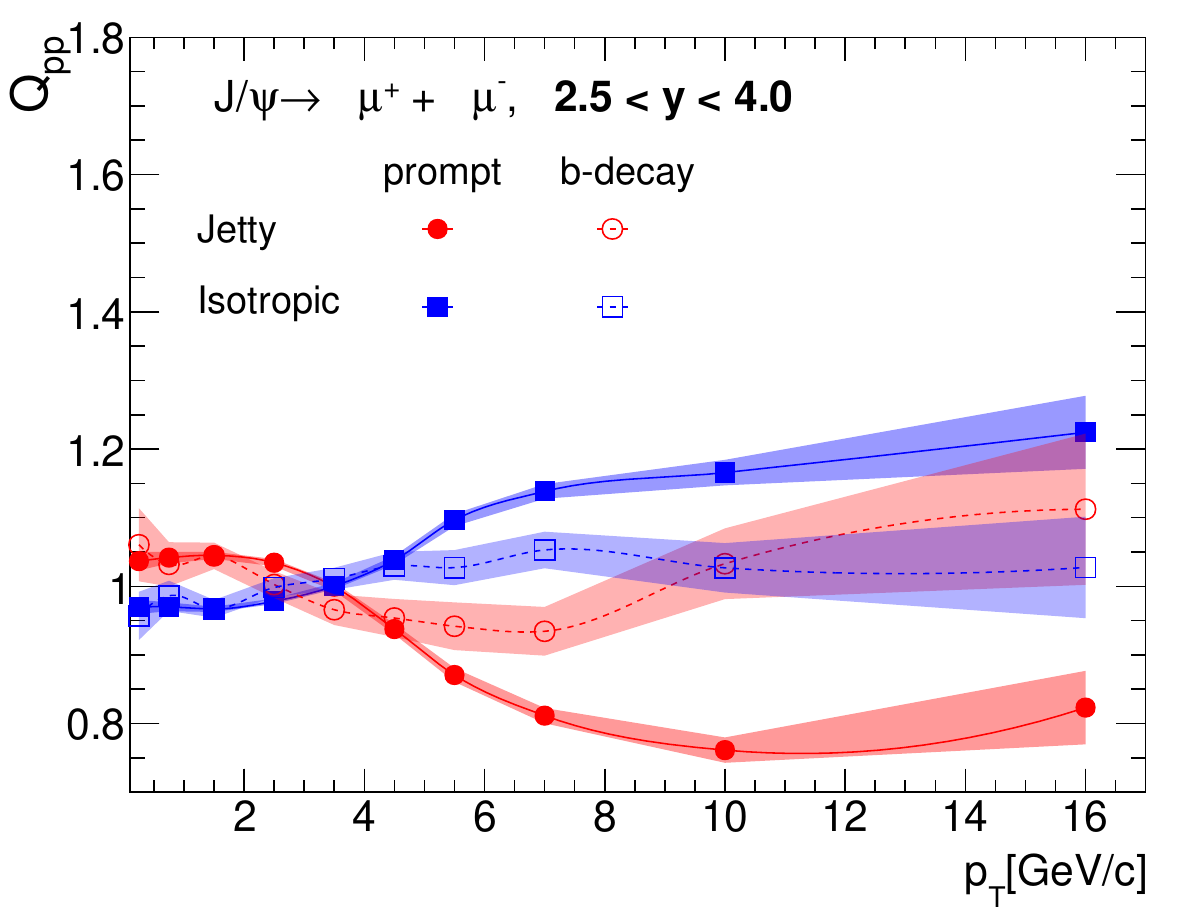}
		\caption{(Color online) $Q_{\rm pp}$ as a function of $p_{\rm T}$ for prompt and non-prompt $\rm{J}/\psi$ in mid (left) and forward (right) rapidities for jetty and isotropic events in $pp$ collisions at $\sqrt{s}$ = 13 TeV using PYTHIA8. Error bands incorporate the estimated statistical uncertainties.}
		\label{fig:Rpp}
	\end{figure*}
	
	A clearer comparison of yields between different $S_{0}$ classes is presented in the bottom panel of Fig.~\ref{fig:pTS0MUEch}, where the ratio of $p_{\rm T}$ spectra for isotropic and jetty events to that of the $S_{0}$-integrated events are shown for both prompt and non-prompt $\rm{J}/\psi$. One can see that the relative prompt $\rm{J}/\psi$ production in isotropic events increases with $p_{\rm T}$ till 10 GeV/$c$ to fall gradually thereafter, indicating the hardness of $p_{\rm T}$-spectra of prompt $\rm{J}/\psi$ in isotropic events in comparison to $S_0$-integrated events. In contrast, it decreases with $p_{\rm T}$ for jetty events to rise after $p_{\rm T}\approx 10$ GeV/$c$, showing the softness of $p_{\rm T}$-spectra in jetty events as compared to $S_0$-integrated events. 
	In Ref.~\cite{Weber:2018ddv}, the authors show that a non-negligible amount of prompt J/$\psi$ are formed through the cluster collapse mechanism, which is largely enhanced for the CR-on scenario for the events with large $N_{\rm mpi}$. Thus, the hardness and softness of $p_{\rm T}$-spectra for the isotropic and jetty events, respectively, for the prompt J/$\psi$ case are a consequence of effects from colour reconnection, where the 
	charm quarks from independent partonic scatterings combine to form charmonium states having a large transverse momentum~\cite{Weber:2018ddv}. The effect of colour reconnection is enhanced for the events having large $N_{\rm mpi}$, consequently, the isotropic event, leading to the observed hardening of prompt J/$\psi$ spectra. At the same time, the yield in both isotropic and jetty events relative to $S_{0}$-integrated events of non-prompt $\rm{J}/\psi$ shows only modest dependence on $p_{\rm T}$ till $p_{\rm T}\approx 6$ GeV/$c$ in both isotropic and jetty events. 
	However, after $p_{\rm T}>6$ GeV/$c$, the relative non-prompt $\rm{J}/\psi$ production abruptly decreases for isotropic events and steeply increases for jetty events to converge with the $S_{0}$-integrated yield. The weaker hardening of non-prompt J/$\psi$ spectra as compared to the prompt case is because of the large mass of the beauty hadrons, which are produced in the initial hard scatterings, which remain unaffected by CR~\cite{Deb:2018qsl}. The effects of hardening (and softening) of $p_{\rm T}$-spectra of prompt and non-prompt J/$\psi$ are diminished at the forward rapidity regions because the effects of MPI are found to be smaller when observations are made in forward rapidity regions, as compared to mid-rapidity, where the particle yield is significantly higher~\cite{MenonKavumpadikkalRadhakrishnan:2023cik}.

\begin{figure*}[ht!]
		\includegraphics[scale=0.43]{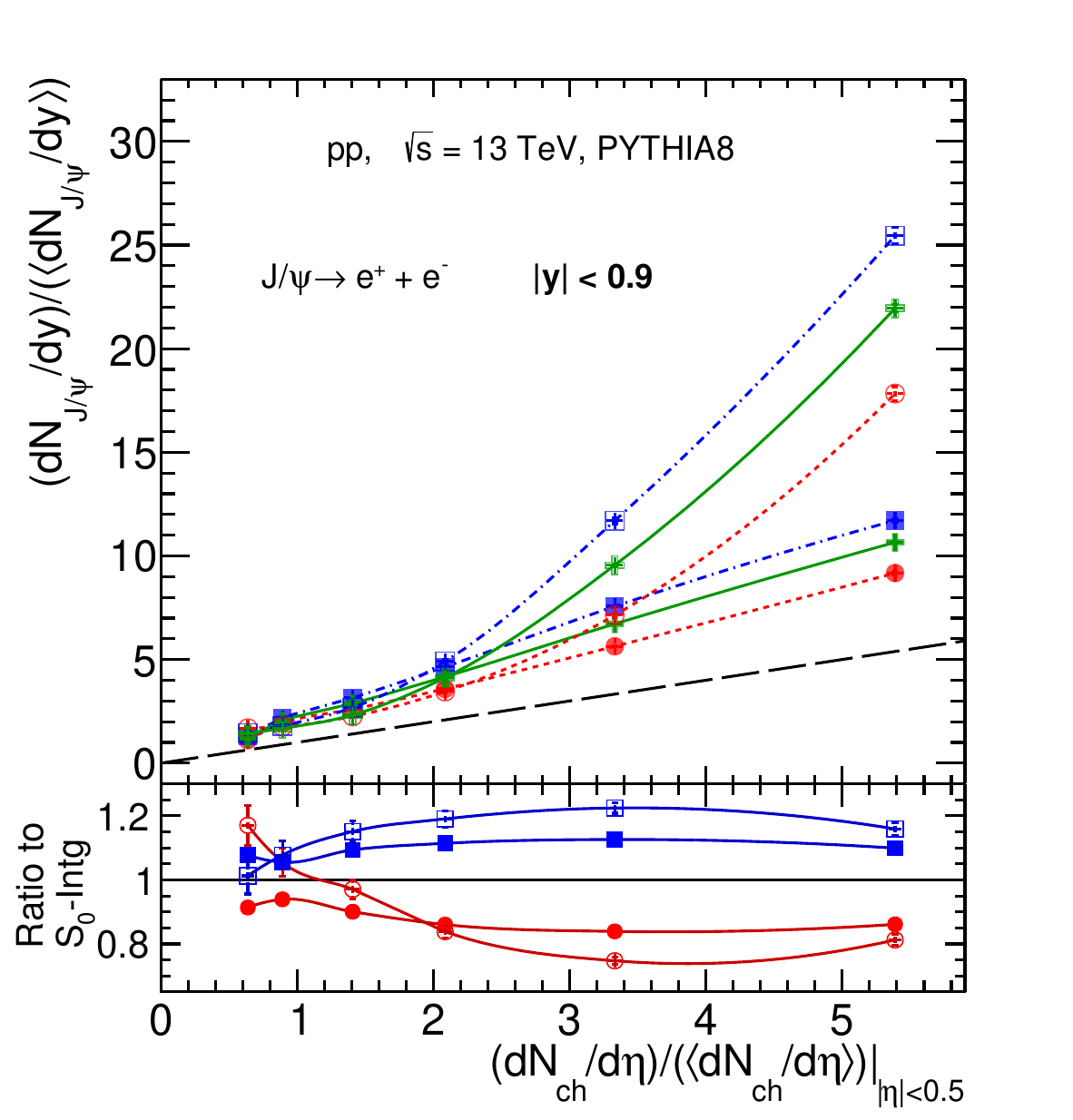}
		\includegraphics[scale=0.43]{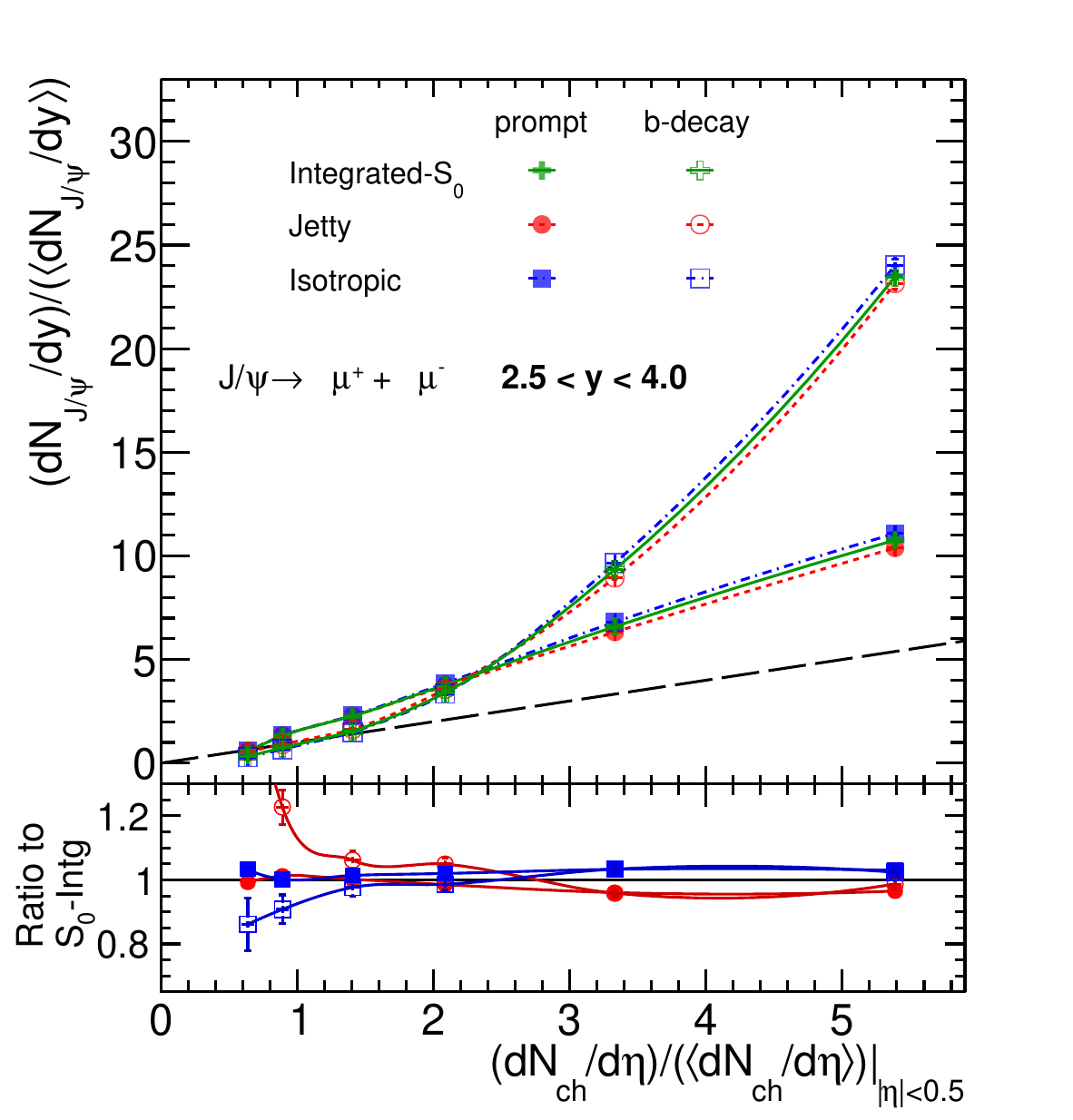}
		\caption{(Color online) Normalized $p_{\rm T}$-integrated prompt and non-prompt $\rm{J}/\psi$ yield as a function of normalized charged particle multiplicity density at midrapidity region ($|\eta| < 0.5$) with V0 multiplicity selection, for $S_{0}$-integrated, jetty and isotropic events from $pp$ collisions at $\sqrt{s}$ = 13 TeV using PYTHIA8, for dielectron (left) and dimuon (right) production channels at $|y| < 0.9$ and $2.5 < y < 4$ rapidity ranges respectively. The dashed black line guides $(dN_{\rm{J}/\psi}/dy)/(\langle dN_{\rm{J}/\psi}/dy \rangle)$ = $(dN_{\rm ch}/d\eta)/(\langle dN_{\rm ch}/d\eta \rangle)$. Estimated statistical uncertainties are well within the marker sizes.}
		\label{fig:NormlzdMUEch}
	\end{figure*}

	In experiments, it is observed that the self-normalized yield of charged particles for $p_{\rm T}>4$ GeV/$c$ shows a stronger than linear increase with an increase in charged particle multiplicity density in the midrapidity region~\cite{ALICE:2019dfi}. These observations are also made for strange~\cite{ALICE:2019avo} and charm~\cite{ALICE:2012pet, ALICE:2020msa, ALICE:2021zkd, ALICE:2025fzz} hadrons at high-$p_{\rm T}$ regions. With PYTHIA8, the stronger-than-linear increase of the self-normalized particle yield of prompt and non-prompt J/$\psi$ is shown in Fig.~\ref{fig:NormlzdMUEch}, the discussion of which will follow later. As discussed in Section~\ref{sec1}, these effects mostly come as a consequence of autocorrelation bias. Also, event classifiers based on multiplicity can lead to contributions from multi-jet scatterings in the final state~\cite{ALICE:2022qxg, ALICE:2023plt}. On the other hand, event classifiers based on event shapes, such as transverse spherocity, can probe $N_{\rm mpi}$\footnote{$S_{0}$ and $N_{\rm ch}$ have a fair degree of correlation with $N_{\rm mpi}$, implying that events having higher $\langle N_{\rm mpi}\rangle$ produces higher final state multiplicity involving softer isotropic particle production~\cite{Prasad:2025yfj}.}, and the contributions from multi-jet topologies can be effectively reduced~\cite{ALICE:2025fzz}. Since these effects have $p_{\rm T}$ dependence, one needs to make a comparison of transverse momentum spectra by taking out the contributions from the final state multiplicity.
	
	\begin{figure*}[ht!]
		\includegraphics[scale=0.43]{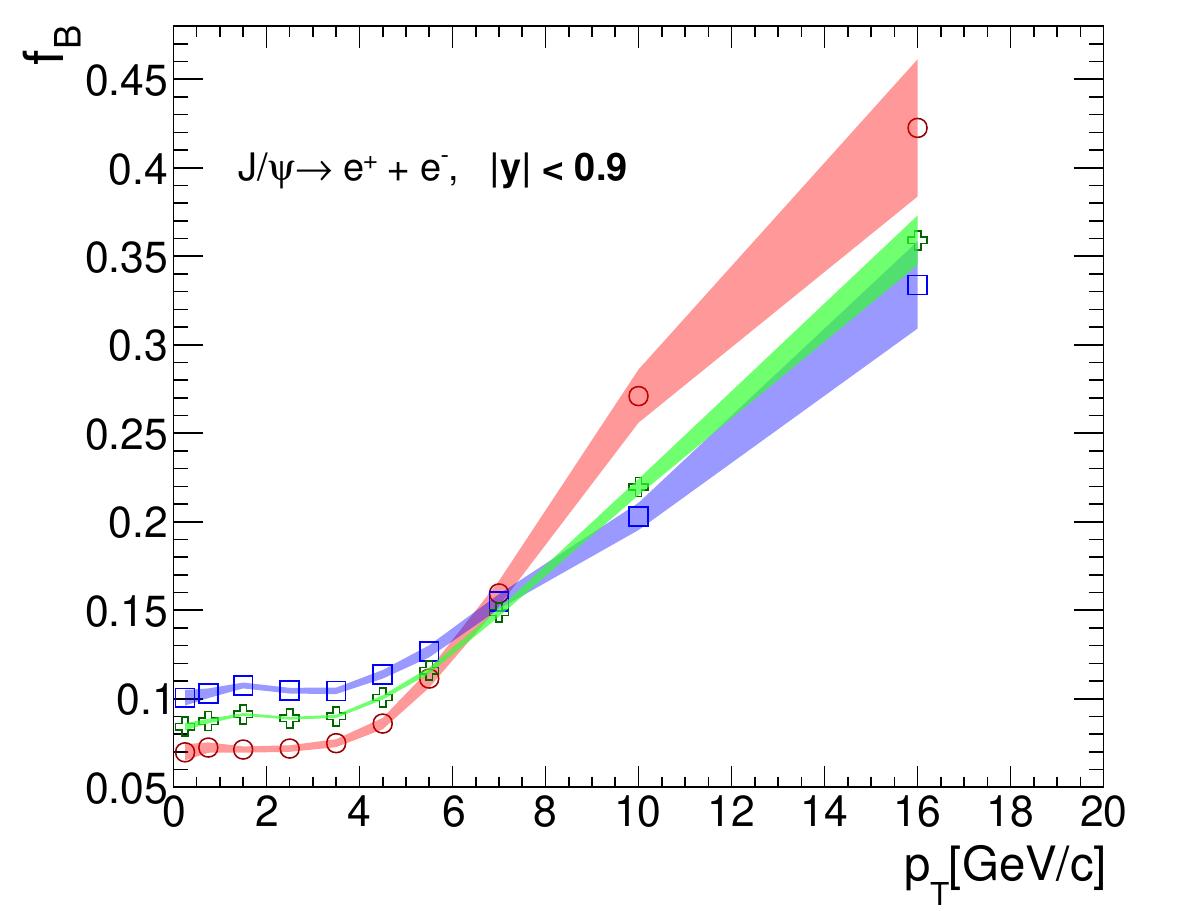}
		\includegraphics[scale=0.43]{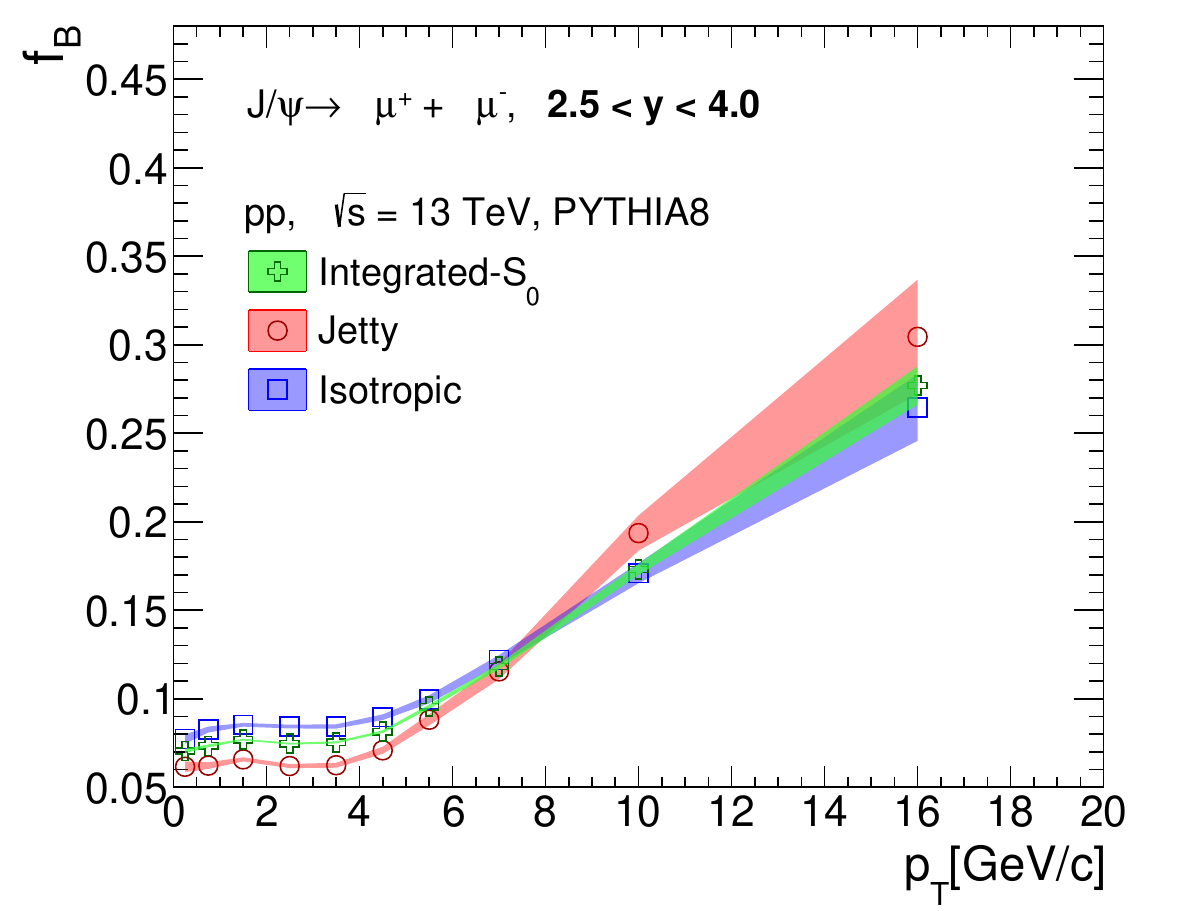}
		\caption{(Color online) The fraction of non-prompt $\rm{J}/\psi$ produced as a function of $p_{\rm T}$ in midrapidity (left) and forward rapidity (right) for $S_{0}$-integrated, jetty and isotropic events from $pp$ collisions at $\sqrt{s}$ = 13 TeV using PYTHIA8. Error bands incorporate the estimated statistical uncertainties.}
		\label{fig:fBMUEch}
	\end{figure*}

	To exclusively understand the modification in the $p_{\rm T}$ spectral shape for prompt and non-prompt J/$\psi$ for isotropic and jetty events, we study the partonic modification factor $Q_{\rm pp}$ whose formulation is inspired by the nuclear modification factor, $R_{\rm AA}$~\cite{ALICE:2019hno}. Through $Q_{\rm pp}$, we aim to make a qualitative comparison of the event-normalized $p_{\rm T}$-differential yield of prompt and non-prompt J/$\psi$ from the isotropic and jetty events to that of $S_{0}$-integrated events, where the yield is scaled by the average number of prompt (or non-prompt) J/$\psi$ in that particular spherocity class. Consequently, $Q_{\rm pp}$ is defined as follows~\cite{MenonKavumpadikkalRadhakrishnan:2023cik, Ortiz:2020rwg}. 
	\begin{equation}
		Q_{\rm pp} = \frac{d^2N_{\rm \rm J/\psi}^{S_{\rm 0}}/\langle N_{\rm \rm J/\psi}^{S_{\rm 0}} \rangle dy dp_{\rm T}} {d^2N_{\rm \rm J/\psi}^{S_{\rm 0}-{\rm int}}/\langle N_{\rm \rm J/\psi}^{S_{\rm 0}{\rm-int}} \rangle dy dp_{\rm T}}
		\label{eq:Rpp}
	\end{equation}
	Here, $\langle N_{\rm \rm J/\psi}^{S_{0}}\rangle$ and $\langle N_{\rm \rm J/\psi}^{S_{\rm 0}-{\rm int}}\rangle$ are the average number of prompt or non-prompt $\rm{J}/\psi$ in corresponding spherocity event class and $S_{0}$-integrated events, respectively. 
	
	Figure~\ref{fig:Rpp} shows the transverse momentum dependence of the partonic modification factor defined in Eq.~\eqref{eq:Rpp} for prompt and non-prompt J/$\psi$. The left and the right panels show the results for the prompt and non-prompt $\rm{J}/\psi$, at mid and forward rapidity regions, respectively. At the mid-rapidity, the trend for both prompt and non-prompt J/$\psi$ is such that it first increases(decreases) with an increase in $p_{\rm T}$ for the isotropic(jetty) case and then starts to fall(rise) towards higher $p_{\rm T}$. However, $Q_{\rm pp}$ attains the peak(valley) at a much earlier $p_{\rm T}$ ($\sim$ 5 GeV/$c$) in the case of non-prompt J/$\psi$ than for the prompt J/$\psi$.
	As discussed earlier, the $p_{\rm T}$ spectra of both prompt and non-prompt J/$\psi$ show different levels of hardening for different classes of spherocity, which is reflected in the positions of peaks and valleys in the $Q_{\rm pp}$ spectra of prompt and non-prompt J/$\psi$. Further, the observed deviations from unity at high-$p_{\rm T}$ for the non-prompt J/$\psi$ measured at the midrapidity region for both the isotropic and jetty events are interesting. As noted in Ref.~\cite{Weber:2018ddv}, the beauty hadron production is often accompanied by a high-energy jet in the opposite direction, which makes the events pencil-like. On the contrary, when the events are isotropic, which corresponds to a high $N_{\rm mpi}$, most of the energy is used up in softer particle production, with little beauty hadrons being produced in the initial scatterings~\cite{Weber:2018ddv}. This shows that the events where the beauty hadrons are produced acquire a unique shape, which is picked up by transverse spherocity. In addition, small auto-correlation bias effects may contribute here as $S_{0}\propto N_{\rm ch}$, which can be avoided when measurements are done at different rapidity regions.
	
	Hence, the measurements made at forward rapidity (right plot) are crucial, as negligible autocorrelation bias is expected in this case. Here, $Q_{\rm pp}$ vary smoothly with increase in $p_{\rm T}$. $Q_{\rm pp}$, which starts slightly above (below) 1 for jetty (isotropic) events at low $p_{\rm T}$, suddenly starts to decrease (increase) after 3 GeV/$c$ for both prompt and non-prompt J/$\psi$. $Q_{\rm pp}$ from isotropic and jetty events has its first crossing at around 3-4 GeV/$c$ for both prompt and non-prompt J/$\psi$ and the second crossing at $\sim$ 10~GeV/$c$ for the non-prompt J/$\psi$ case. However, unlike in midrapidity, here at forward rapidity, the non-prompt J/$\psi$ shows no deviations from unity. This is a testimony that the production of beauty hadrons at forward rapidity would not modify the event shape measured through the charged particles in the midrapidity region. Moreover, the $p_{\rm T}$ dependence of $Q_{\rm pp}$ is found to be in alignment within uncertainties when the study is repeated for the CR-BLC~2 PYTHIA model for both the dileptonic decay channels of J/$\psi$ (particularly prompt J/$\psi$).
	
	\begin{figure*}[ht!]
		\includegraphics[scale=0.43]{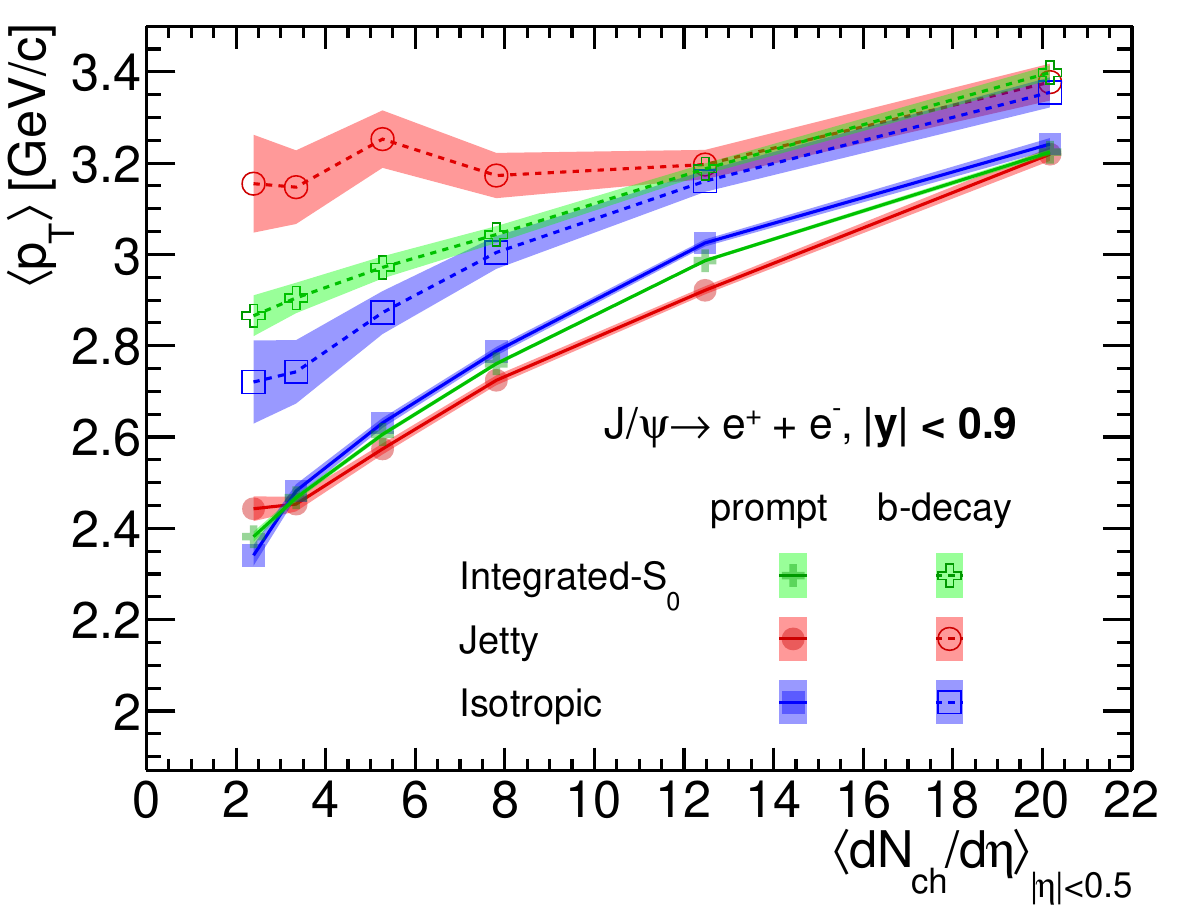}
		\includegraphics[scale=0.43]{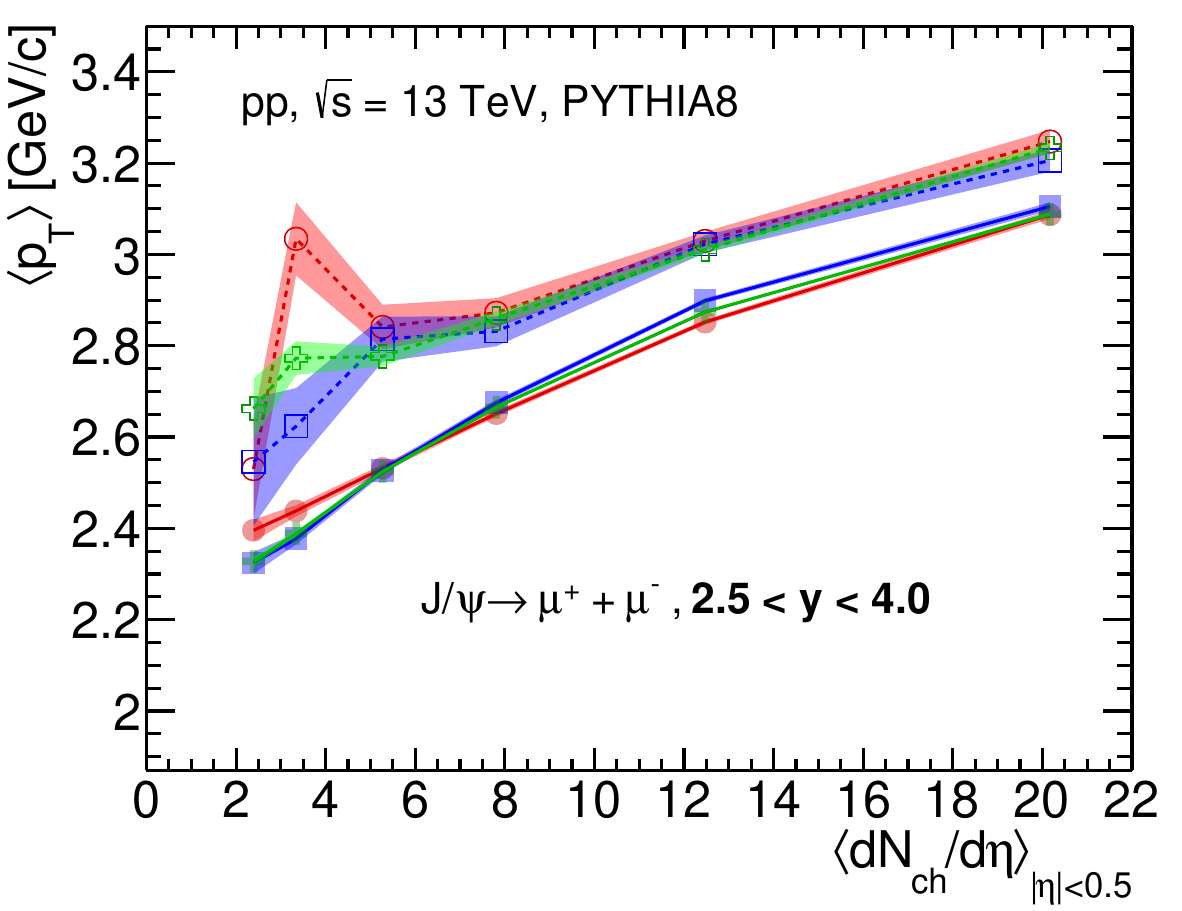}
		\caption{(Color online) Mean transverse momentum ($\langle p_{\rm T} \rangle$) as a function of midrapidity charged particle multiplicity density for prompt and non-prompt $\rm{J}/\psi$ in mid (left) and forward (right) rapidity regions for $S_{0}$-integrated, jetty and isotropic events for $pp$ collisions at $\sqrt{s}$ = 13 TeV using PYTHIA8. Error bands incorporate the estimated statistical uncertainties.}
		\label{fig:MeanpTHIL}
	\end{figure*}

	To understand the contribution of $\rm{J}/\psi$ produced from the decay of $b$-hadrons to the inclusive $\rm{J}/\psi$ production, the fraction of non-prompt to inclusive $\rm{J}/\psi$ ($f_{\rm B}$) is studied as a function of transverse momentum and transverse spherocity in (0-100\%) V0M class, both at midrapidity and forward rapidity regions, as shown in Fig.~\ref{fig:fBMUEch}. A qualitative similarity of our results with the ALICE and LHCb data could be observed from the comparison presented in Fig.~\ref{fig:ALICELHCbfB} in the Appendix~\ref{apxdata}. Similar to the experimental observations at midrapidity in Ref.~\cite{ALICE:2021edd}, the non-prompt $\rm{J}/\psi$ fraction stays independent of $p_{\rm T}$ for $p_{\rm T}\leq4$ GeV/$c$ after which $f_{\rm B}$ increases steadily with an increase in $p_{\rm T}$. This relative increase in the number of non-prompt J/$\psi$ to inclusive J/$\psi$ with an increase in $p_{\rm T}$ is consistent with the observations of hardening of the $p_{\rm T}$-spectra of non-prompt J/$\psi$ with respect to prompt J/$\psi$, as shown in Fig.~\ref{fig:pTS0MUEch}, for both the rapidity regions. In addition, $f_{\rm B}$ shows significant transverse spherocity dependence. Here, at the midrapidity region, for $p_{\rm T}\leq6$ GeV/$c$, isotropic events show a larger value of $f_{\rm B}$. Interestingly, this trend is reversed for $p_{\rm T}>6$ GeV/$c$, where jetty events lead $f_{\rm B}$. This transverse spherocity dependence at the low $p_{\rm T}$ region reflects the effects of MPI on the production of beauty hadrons. In this low $p_{\rm T}$ regime, a larger fraction of non-prompt J/$\psi$ is produced in the isotropic events (large $N_{\rm mpi}$) as compared to the jetty events (having small $N_{\rm mpi}$). This is, hence, an affirmation that, in PYTHIA8, MPI activity can affect the beauty hadron production~\cite{Weber:2018ddv}. However, at high $p_{\rm T}$ regions, a large fraction of beauty hadrons are produced from the initial hard scattering itself. These hard scatterings are significantly large in jetty events by their very nature, which leads to the enhanced value of $f_{\rm B}$ at large values of $p_{\rm T}$. A similar observation is also made at the forward rapidity region. Here, however, the crossing of $f_{\rm B}$ for jetty and isotropic events happens at $p_{\rm T}\approx8$ GeV/$c$. In addition, we find $f_{\rm B}$ to be smaller in forward rapidity as compared to the midrapidity region, for all $S_0$ classes. Note that these differences based on rapidity are also qualitatively in match with the ALICE and LHCb experimental data, provided in Fig.~\ref{fig:ALICELHCbfB} (Appendix~\ref{apxdata}). A comparison with the CR-BLC~2 configuration of PYTHIA8 indicates that the qualitative trend and the segregation power of $S_{0}$ are valid irrespective of model differences, though an average reduction of $\sim40\%$ relative to Tune 4C is observed in both the dielectron and dimuon channels over the studied $p_{\rm T}$ range.

	Figure~\ref{fig:MeanpTHIL} shows the mean transverse momentum ($\langle p_{\rm T}\rangle$) as a function of average charged particle multiplicity at midrapidity ($\langle dN_{\rm ch}/d\eta \rangle_{|\eta|<0.5}$) for prompt and non-prompt $\rm{J}/\psi$ in the three $S_{0}$ classes for measurements at mid rapidity (left plot) and forward rapidity (right plot) in $pp$ collisions at $\sqrt{s}=13$ TeV using PYTHIA8.  $\langle p_{\rm T}\rangle$ of both prompt and non-prompt $\rm{J}/\psi$ is seen to rise with multiplicity, where midrapidity measurements achieve greater magnitudes of $\langle p_{\rm T}\rangle$ than those in forward rapidity. This observation of rapidity dependence of $\langle p_{\rm T}\rangle$ is consistent with the results of light flavour hadrons, as shown in Ref.~\cite{MenonKavumpadikkalRadhakrishnan:2023cik}. For both the pseudorapidity regions, one finds a larger $\langle p_{\rm T}\rangle$ for the non-prompt J/$\psi$ as compared to the prompt case. Since the non-prompt J/$\psi$ carries most of the transverse momentum of its parent beauty hadrons, $\langle p_{\rm T}\rangle$ of non-prompt J/$\psi$ is the reflection of $\langle p_{\rm T}\rangle$ of the decaying beauty hadrons, which is evidently larger than the $\langle p_{\rm T}\rangle$ of prompt J/$\psi$. This is because the beauty hadrons that contribute to non-prompt J/$\psi$ possess a larger mass compared to the charmonium states. In other words, Fig.~\ref{fig:MeanpTHIL} reflects the mass ordering of $\langle p_{\rm T}\rangle$, which is consistent with the mass-ordering of $\langle p_{\rm T}\rangle$ for different hadrons in the light-flavour sector~\cite{ALICE:2019hno}. There is also a notable $S_{0}$ dependence for the enhancement of $\langle p_{\rm T}\rangle$ with increase in $\langle dN_{\rm ch}/d\eta \rangle_{|\eta|<0.5}$, especially for the $\rm{J}/\psi$ reconstructed at midrapidity. At low multiplicity density, $\langle p_{\rm T}\rangle$ of prompt and non-prompt $\rm{J}/\psi$ from jetty events have a slight dominance in magnitude over the $\rm{J}/\psi$ produced from isotropic events. This is because the jetty events of the lowest multiplicity class are strongly dominated by the hard scatterings; the hadrons produced in such interactions have large transverse momenta, consequently possessing a large value of $\langle p_{\rm T}\rangle$ as compared to isotropic events in a similar multiplicity class. This difference is clearer in the case of midrapidity non-prompt $\rm{J}/\psi$, the production source of which is primarily the initial hard scatterings. Here, at higher multiplicities ($\langle dN_{\rm ch}/d\eta\rangle _{|\eta|<0.5}\gtrsim 12$), the distinction based on spherocity diminishes for non-prompt $\rm{J}/\psi$ while prompt $\rm{J}/\psi$ from isotropic events possess slightly higher $\langle p_{\rm T}\rangle$ than their counterparts from jetty events. This indirectly signifies the role of MPI in charm hadron production in contrast to beauty hadrons. Further, one finds that the spherocity dependence of the $\langle p_{\rm T}\rangle$ of prompt and non-prompt J/$\psi$ is diminished at the forward rapidity.

	The upper panels in Fig.~\ref{fig:NormlzdMUEch} show the self-normalized yield of prompt and non-prompt $\rm{J}/\psi$ as a function of self-normalized charged particle multiplicity density at mid-pseudorapidity ($|\eta|<0.5$), for $S_{0}$-integrated, jetty and isotropic events in $pp$ collisions at $\sqrt{s}=13$ TeV using PYTHIA8. The left and right panels show the results at the mid and forward rapidities, respectively. The normalized $\rm{J}/\psi$ yield for both prompt and non-prompt cases increases with normalized charged particle density for measurements at both mid and forward rapidities and for all classes of transverse spherocity. However, in both the rapidity regions, for the $S_{0}$-integrated case, one finds that the increment rate of non-prompt J/$\psi$ is stronger as compared to the prompt case with an increase in self-normalized charged particle density. This is because the beauty hadrons are usually formed in initial hard scattering and are accompanied by high-energy jets, which can further fragment to produce hadrons, which can increase the charged particle multiplicity in the final state~\cite{Weber:2018ddv}. Thus, the stronger-than-linear increase in non-prompt J/$\psi$ production can be attributed to an autocorrelation bias~\cite{Weber:2018ddv,ALICE:2025fzz}. As the particle multiplicity selection is performed in the forward rapidity region, one finds a comparatively smaller autocorrelation bias in the left panel, where both prompt and non-prompt J/$\psi$ are measured in the midrapidity region, in contrast to the right panel, where the J/$\psi$ is measured in the forward rapidity region.
	
	Furthermore, the effect of this autocorrelation bias can be further explored with transverse spherocity. In Ref.~\cite{Weber:2018ddv}, the influence of $N_{\rm mpi}$ is explored for both inclusive and non-prompt J/$\psi$. Here, the effects of $N_{\rm mpi}$ are similar for both inclusive and non-prompt J/$\psi$, except for extreme $N_{\rm mpi}$ events, where the yield of non-prompt J/$\psi$ saturates. Interestingly, in the left panel of Fig.~\ref{fig:NormlzdMUEch}, we find that the yield of both prompt and non-prompt J/$\psi$ is enhanced for the isotropic events having large $N_{\rm mpi}$, while a lower self-normalized yield is observed for the jetty events with a small $N_{\rm mpi}$ value. This is a contradiction to the above-mentioned dependence of non-prompt J/$\psi$ production with $N_{\rm mpi}$, and can be attributed to self-correlation bias caused by measuring transverse spherocity in a similar rapidity region as the J/$\psi$. Although this bias is present in both prompt and non-prompt J/$\psi$ measured at midrapidity, its production mechanism enhances the effects for the non-prompt case, which can be clearly visible in the lower left panel of Fig.~\ref{fig:NormlzdMUEch}. In contrast, the dependence of transverse spherocity is small on both prompt and non-prompt J/$\psi$ measured at the forward rapidity, where the $S_0$ induced bias is small. Additionally, the efficiency of $S_{0}$ in categorizing J/$\psi$ production topologically is found to be maintained when the results are calculated using the CR-BLC~2 settings of PYTHIA8. Moreover, the increase of self-normalised yield of prompt and non-prompt $\rm{J}/\psi$ with respect to multiplicity from the CR-BLC~2 model is observed to very closely follow the trend from Tune 4C of PYTHIA8, such that the magnitude of the former is $\sim(10-15)\%$($\sim(15-25)\%$) lesser than the latter in the highest (lowest) multiplicity class considered.

	\section{Summary}
	\label{sec:summary}
	In this work, we study the transverse spherocity dependence of prompt and non-prompt J/$\psi$ production in $pp$ collisions at $\sqrt{s}=13$ TeV using PYTHIA8. For the first time, the topological production of J/$\psi$ is studied with an event shape. The study shows the effect of bias and the effects of MPI by studying the topological production of J/$\psi$ in different pseudorapidity regions. We find that the non-prompt J/$\psi$ possesses a harder transverse momentum spectrum as compared to the prompt J/$\psi$. The prompt J/$\psi$ spectra are harder for the isotropic events as compared to the jetty events. The hardness of the $p_{\rm T}$-spectra for the isotropic events, which may arise as a consequence of multi-jet production environments and events with significantly large underlying event activity, in contrast to jetty events where the production of J/$\psi$ is mostly dominated by di-jet like events, is noteworthy. This suggests that, although the J/$\psi$ is produced in the initial stages, underlying MPI plays a crucial role in shaping the $p_{\rm T}$-spectra. The effect extends to the beauty hadron production. A similar observation is made in Ref.~\cite{CMS:2020fae} for $\Upsilon(\rm nS)$ production. 
	
	Moreover, the non-prompt production fraction for the jetty events, dominated by the di-jet structure of particle production, increases with $p_{\rm T}$ and becomes higher than that of isotropic events. This identifies the dominance of the production of beauty hadron fraction in di-jet events in contrast to events with multi-jet environments. With enough statistics in Run 3 of LHC data taking, this property of transverse spherocity can be exploited to study the production of charmonia in jets and offer valuable insights into the underlying hadronization mechanisms from pp to Pb–Pb collisions. This study provides a crucial understanding of charm and beauty hadron production through the study of prompt and non-prompt J/$\psi$ with respect to different event shapes, including di-jet and multi-jet environments, where transverse spherocity acts as a probe.
	
	Further, the self-normalized yield of J/$\psi$ in the forward rapidity as a function of self-normalized charged particle multiplicity is affected by a higher autocorrelation bias as compared to the measurement of J/$\psi$ in the midrapidity as the multiplicity selection is performed in the forward rapidity regions. In contrast, this autocorrelation bias is enhanced in the mid-rapidity region for transverse spherocity selection, which is defined in the mid-rapidity region. This warns the scientific community to be careful while dealing with particle production at different rapidity regions and while making physics conclusions out of it.
	
	\appendix
	\section*{Appendix}
	\subsection{Comparison with experimental data}
	\label{apxdata}
	
\begin{figure*}[ht!]
		\includegraphics[scale=0.4]{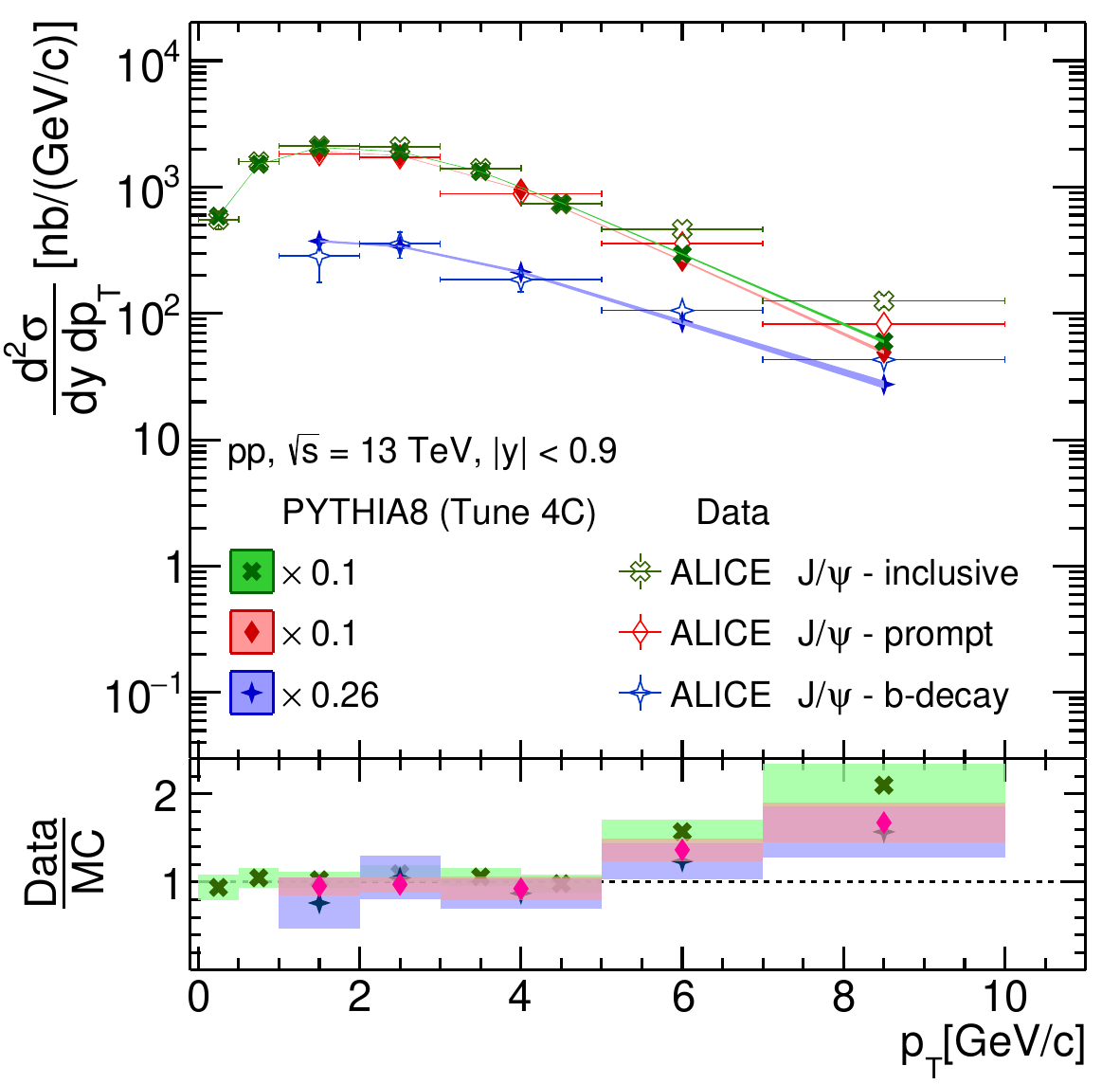}
		\includegraphics[scale=0.4]{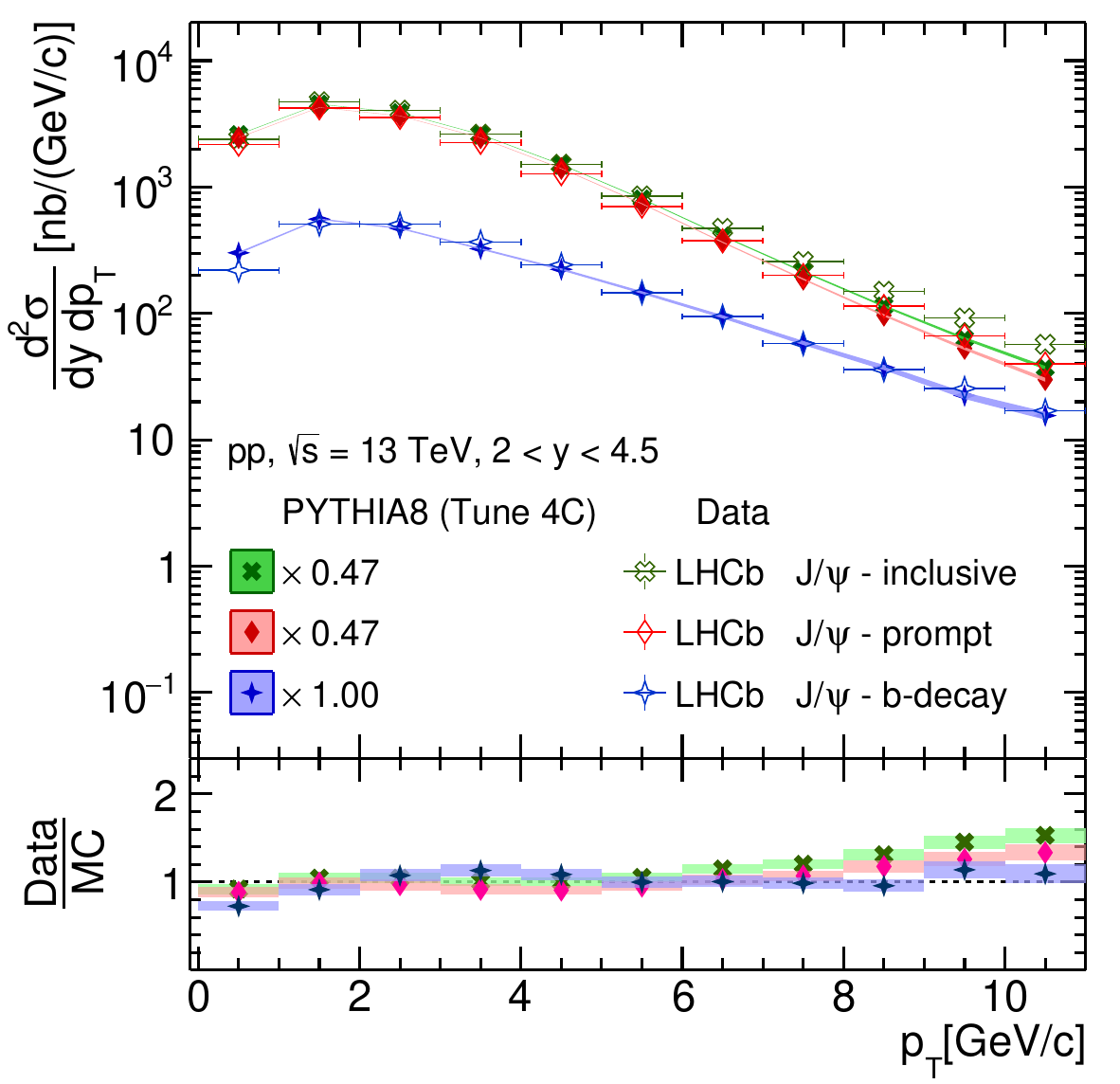}
		\caption{(Color online) Comparison of inclusive, prompt and non-prompt $\rm{J}/\psi$ production cross-section from PYTHIA8 (Tune~4C) with ALICE data~\cite{ALICE:2021dtt, ALICE:2021edd} at midrapidity (left) and with LHCb data~\cite{LHCb:2015foc} at forward rapidity (right) for $pp$ collisions at $\sqrt{s}$~=~13~TeV. The bottom panel shows the ratio of experimental data to the PYTHIA8 Tune 4C results.}
		\label{fig:Sigma3JpsiMUEch}
	\end{figure*}

	In Fig.~\ref{fig:Sigma3JpsiMUEch}, we compare the production cross-section of inclusive, prompt and non-prompt $\rm{J}/\psi$ from PYTHIA8 simulation of $pp$ collisions at $\sqrt{s}$ = 13 TeV with the corresponding data from ALICE~\cite{ALICE:2021dtt, ALICE:2021edd} and LHCb~\cite{LHCb:2015foc} for the midrapidity $\rm{J}/\psi$ (left plot) and forward rapidity $\rm{J}/\psi$ (right plot) respectively. Similar to the observation in Ref.~\cite{Prasad:2023zdd} and consistent with Fig.~\ref{fig:pTS0MUEch}, the production yield of $\rm{J}/\psi$ from $b$-hadron decays is almost 10 times smaller than the prompt $\rm{J}/\psi$ for the lower $p_{\rm T}$ values while this difference gets reduced at higher values of $p_{\rm T}$. In the case of $\rm{J}/\psi$ reconstructed at midrapidity, the results from PYTHIA8 overestimate the experimental ALICE data. Thus, scaling factors of 0.1, 0.1, and 0.26 are applied respectively to the PYTHIA8 data of inclusive, prompt, and non-prompt $\rm{J}/\psi$. Similarly, for the $\rm{J}/\psi$ measurements at forward rapidity, we multiply the PYTHIA8 results of inclusive and prompt $\rm{J}/\psi$ by 0.47 to match the experimental data from LHCb. We see that the results from PYTHIA8 follow the experimental results fairly well till $p_{\rm T} <$ 6 TeV in both mid and forward rapidities, to deviate thereafter for the higher values of $p_{\rm T}$.

	\begin{figure}
		\includegraphics[scale=0.4]{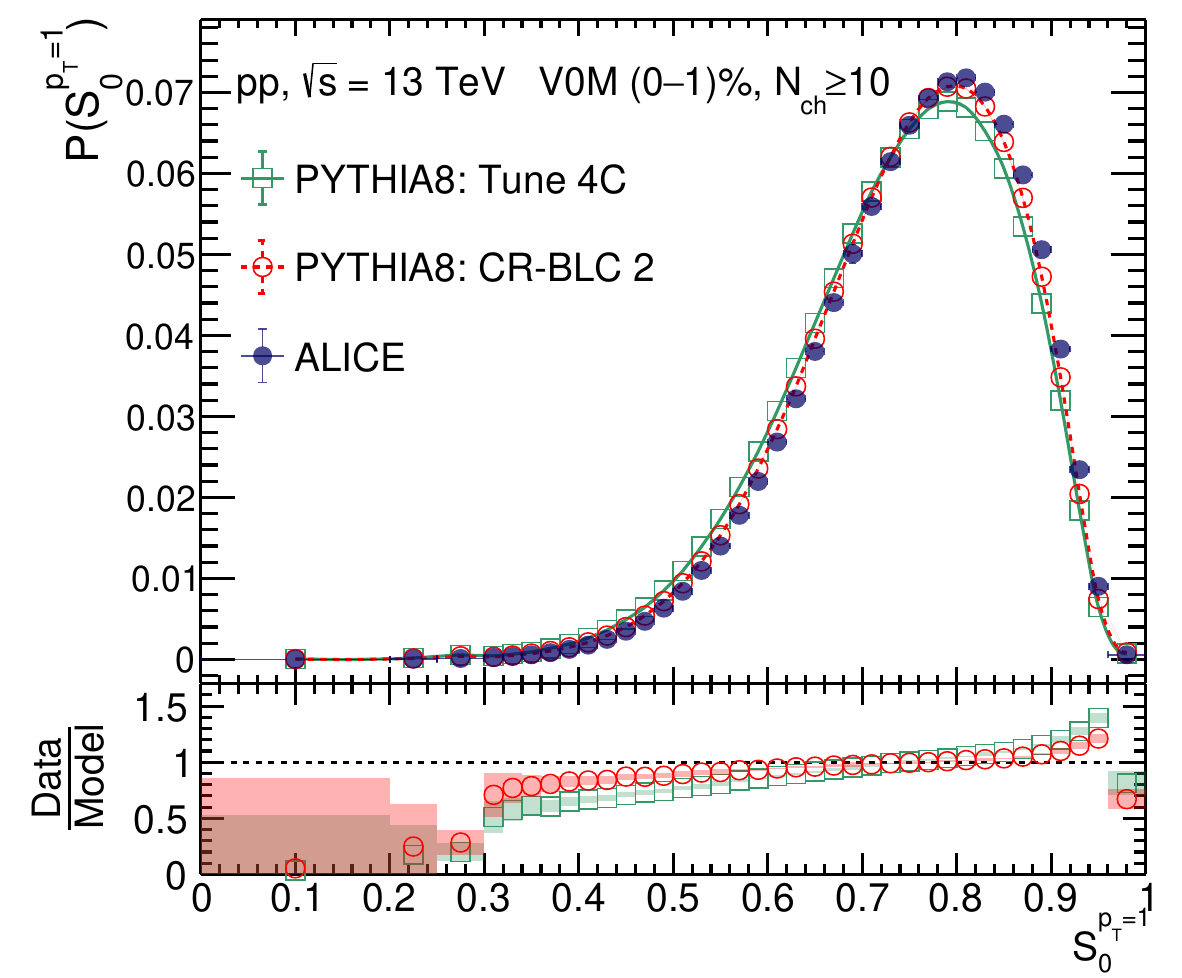}
		\caption{Event normalized $S_{0}^{p_{\rm T}=1}$ distribution for the 0--1\% V0M multiplicity class for $pp$ collisions at $\sqrt{s}_{\rm}=13$~TeV from PYTHIA8 (Tune 4C and CR-BLC~2 settings) and the experimental data from ALICE~\cite{ALICE:2023bga} (upper panel). Ratio in the lower panel compares ALICE experimental data to PYTHIA8 results.}
		\label{fig:S0datacomp}
	\end{figure}
	
	\begin{figure}
		\includegraphics[scale=0.4]{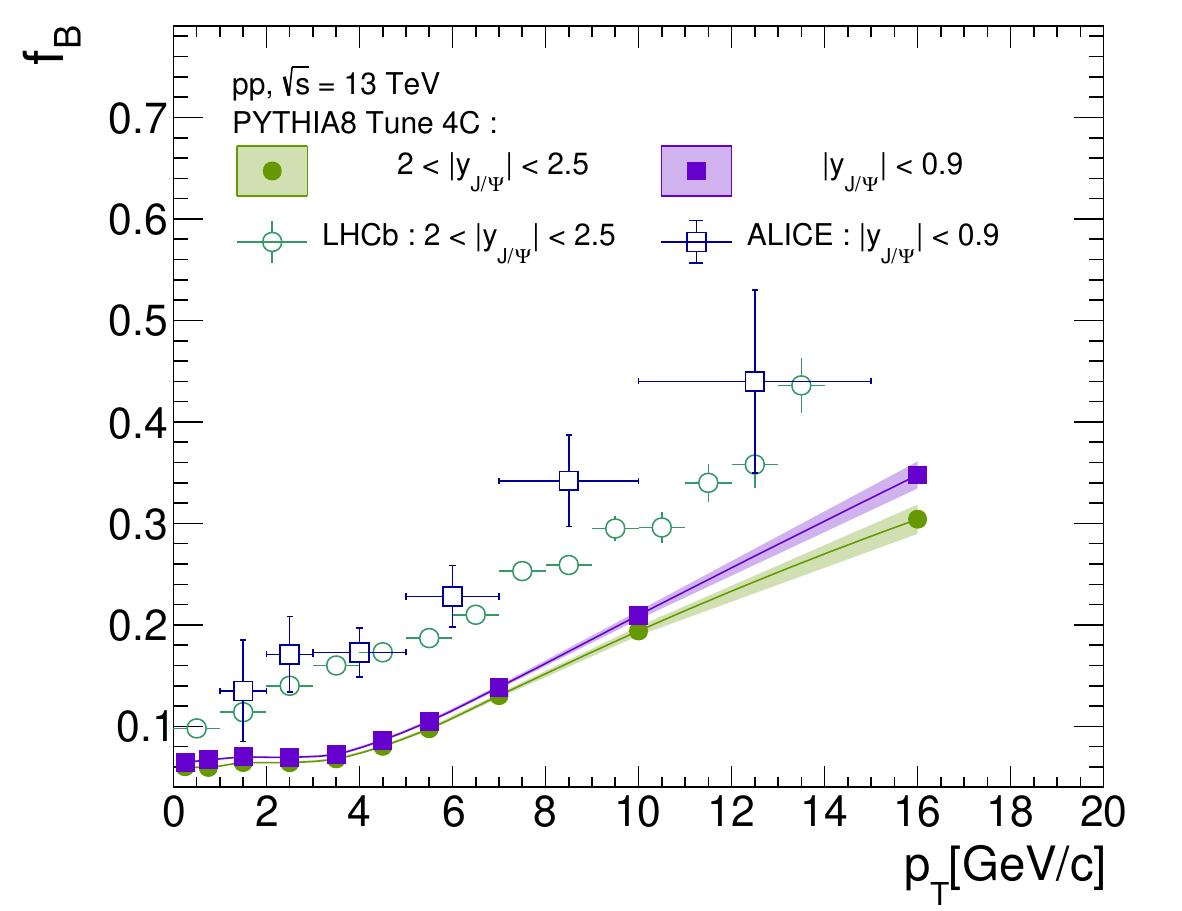}
		\caption{Comparison of fraction of non-prompt $\rm{J}/\psi$ produced as a function of $p_{\rm T}$ for minimum-bias events from $pp$ collisions at $\sqrt{s}$ = 13 TeV from PYTHIA8 (Tune 4C) with ALICE (midrapidity) and LHCb (forward rapidity) data ~\cite{ALICE:2021edd, LHCb:2015foc}.}
		\label{fig:ALICELHCbfB}
	\end{figure}
	
	\subsection{Model Comparisons}
	\label{apxmodel}
	
\begin{figure*}
		\includegraphics[scale=0.445]{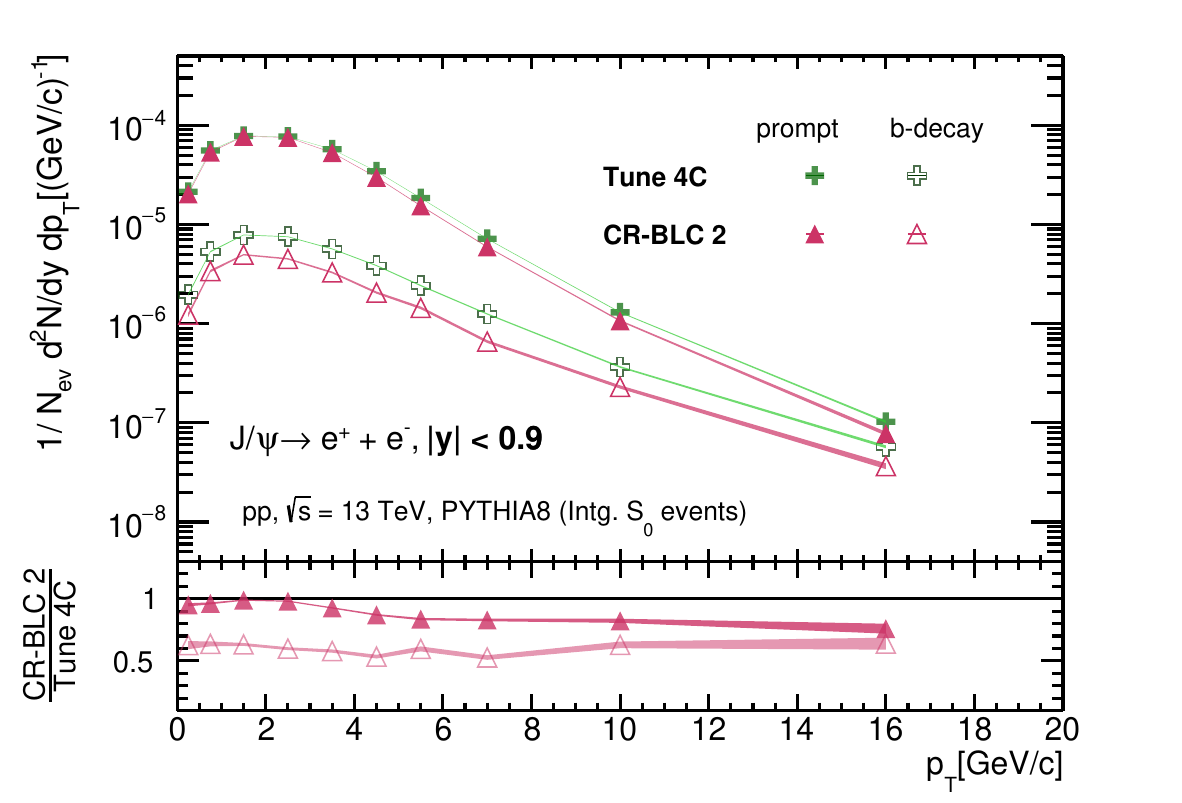}
		\includegraphics[scale=0.445]{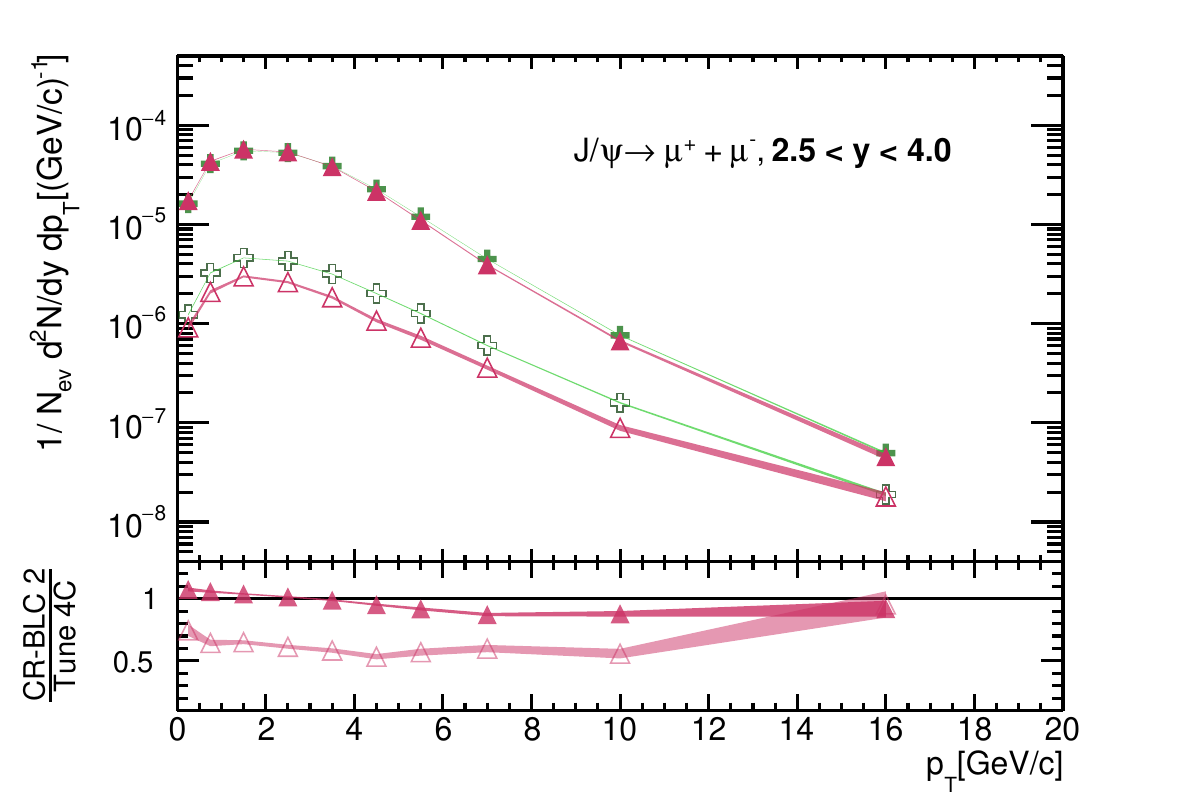}
		\includegraphics[scale=0.88]{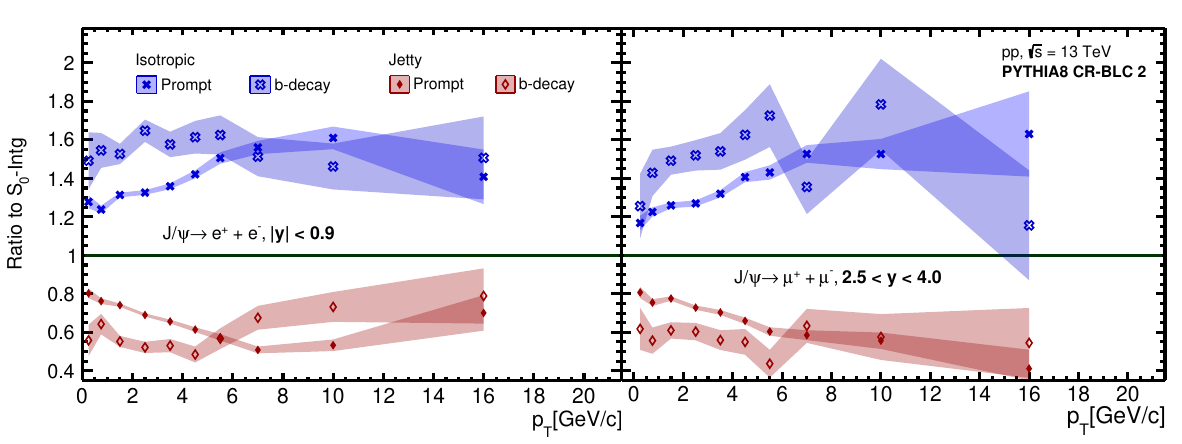}
		
		\caption{Comparison of the $p_{\rm T}$ spectra of prompt and non-prompt $\rm{J}/\psi$ between Tune 4C and CR-BLC~2 settings of PYTHIA8 for $pp$ collisions at $\sqrt{s}$ = 13 TeV for the integrated spherocity events is shown in the upper plots. The ratio of $p_{\rm T}$ spectra of the jetty and isotropic to $S_{0}$-integrated events for prompt and non-prompt $\rm{J}/\psi$ for the PYTHIA8 CR-BLC~2 setting is shown in the bottom panel. Error bands incorporate the estimated statistical uncertainties.}
		\label{fig:CRBLCvs4C}
	\end{figure*}

	To assess the robustness of the results against modelling uncertainties, the analysis was repeated using the PYTHIA CR-BLC~2 colour reconnection model, built upon the Monash 2013 tune (hereafter referred to as CR-BLC~2). This model is known to introduce beyond-leading-colour effects, including junction formation and time-dependent reconnections, which differ significantly from the simplified colour reconnection scheme used in Tune 4C~\cite{ALICE:2023brx, Christiansen:2015yqa}. As can be seen in the upper row of Fig.~\ref{fig:CRBLCvs4C}, the $\rm{J}/\psi$ yield with respect to $p_{\rm T}$ for the $S_{0}$-integrated events from \textit{pp} collisions at $\sqrt{s}_{\rm}=13$~TeV for the CR-BLC~2 model is (10-20)$\%$ lower relative to Tune 4C, for prompt $\rm{J}/\psi$, and for both $\rm{J}/\psi$ decay channels. This difference increases as we shift our focus to non-prompt $\rm{J}/\psi$. Although both configurations employ PYTHIA8 as the event generator, they incorporate fundamentally different treatments of non-perturbative QCD dynamics, especially in the hadronisation stage. In particular, the inclusion of junction formation and time-dilated colour reconnections in the CR-BLC~2 model leads to a redistribution of colour flux and energy during the hadronisation stage, which can affect the neutralisation of heavy quark pairs and, consequently, the formation probability of bound charmonium states~\cite{Christiansen:2015yqa}. This yield reduction is more visible for the non-prompt $\rm{J}/\psi$, possibly due to the enhanced sensitivity of beauty-hadron hadronisation to junction-rich colour reconnection effects. Such effects are absent in the simpler string-length–minimization-based colour reconnection model employed in Tune 4C~\cite{Christiansen:2015yqa}. 
	
	However, the actual important factor is the qualitative similarity one observes between the trends of $p_{\rm T}$ spectra from both PYTHIA settings. The lower plot in Fig.~\ref{fig:CRBLCvs4C} shows the ratios of $p_{\rm T}$ spectra from jetty (or isotropic) events to that of $S_{0}$-integrated events, for prompt and non-prompt $\rm{J}/\psi$, for both the decay channels of $\rm{J}/\psi$ from PYTHIA CR-BLC~2 model. The trends can be seen once again to be similar to the lower panel of Fig.~\ref{fig:pTS0MUEch}, which corresponds to PYTHIA8 Tune 4C, which is the default setting employed for event generation throughout this work.
	Apart from these, the trends of all the major observables studied using PYTHIA8 Tune 4C are seen to be qualitatively replicated by the CR-BLC~2 model. The $S_{0}$, $p_{\rm T}$ and self-normalised $dN_{\rm ch}/d\eta$ dependence of all these observables are observed to be well-preserved despite the PYTHIA tuning/setting differences.

	\section*{Acknowledgement}
	A.M.K.R. acknowledges the doctoral fellowships from the DST INSPIRE program of the Government of India. N.M. is supported by the Academy of Finland through the Center of Excellence in Quark Matter with Grant No. 346328. The authors gratefully acknowledge the DAE-DST, Government of India, funding under the mega-science project “Indian participation in the ALICE experiment at CERN” bearing Project No. SR/MF/PS-02/2021-IITI(E-37123).

\end{document}